\documentclass[sigconf]{aamas} 
\renewcommand\footnotetextcopyrightpermission[1]{} 
\usepackage{balance} 

\usepackage{booktabs} 
\usepackage{tikz} 
\newcommand*\circled[1]{\tikz[baseline=(char.base)]{
            \node[shape=circle,draw,inner sep=2pt] (char) {#1};}}

\usepackage{algorithm}
\usepackage{algpseudocode}
\usepackage{xcolor}
\usepackage{booktabs}
\usepackage{enumitem}
\usepackage{url}
\usepackage{multicol}
\usepackage{multirow}
\usepackage{arydshln}
\usepackage{adjustbox}
\usepackage{soul}
\usepackage{bbm}

\renewcommand{\paragraph}[1]{\smallskip\noindent\textbf{#1}}
\newenvironment{rcases}
  {\left.\begin{aligned}}
  {\end{aligned}\right\rbrace}
\usepackage{pifont}
\setcopyright{ifaamas}
\acmConference[AAMAS '24]{Proc.\@ of the 23rd International Conference
on Autonomous Agents and Multiagent Systems (AAMAS 2024)}{May 6 -- 10, 2024}
{Auckland, New Zealand}{N.~Alechina, V.~Dignum, M.~Dastani, J.S.~Sichman (eds.)}
\copyrightyear{2024}
\acmYear{2024}
\acmDOI{}
\acmPrice{}
\acmISBN{}

\usepackage[absolute,overlay]{textpos}
\title[Achieving Fairness in Transaction Fee Mechanism Design]{No Transaction Fees? No Problem! Achieving Fairness in Transaction Fee Mechanism Design}

\author{Sankarshan Damle}
\affiliation{
  \institution{IIIT, Hyderabad}
  \city{Hyderbad}
  \country{India}}
\email{sankarshan.damle@research.iiit.ac.in}

\author{Varul Srivastava}
\affiliation{
  \institution{IIIT, Hyderabad}
  \city{Hyderbad}
  \country{India}}
\email{varul.srivastava@research.iiit.ac.in	}

\author{Sujit Gujar}
\affiliation{
  \institution{IIIT, Hyderabad}
  \city{Hyderbad}
  \country{India}}
\email{sujit.gujar@iiit.ac.in}

\begin{abstract}
The recently proposed Transaction Fee Mechanism (TFM) literature studies the strategic interaction between the miner of a block and the transaction creators (or users) in a blockchain. In a TFM, the miner includes transactions that maximize its utility while users submit fees for a slot in the block. The existing TFM literature focuses on satisfying standard incentive properties -- which may limit widespread adoption. We argue that a TFM is ``fair" to the transaction creators if it satisfies specific notions, namely Zero-fee Transaction Inclusion and Monotonicity. First, we prove that one generally cannot ensure both these properties and prevent a miner's strategic manipulation. We also show that existing TFMs either do not satisfy these notions or do so at a high cost to the miners' utility. As such, we introduce a novel TFM using on-chain randomness -- \rftm. We prove that \rftm\ guarantees incentive compatibility for miners and users while satisfying our novel fairness constraints.
\end{abstract}

\keywords{Transaction Fee Mechanism Design, Fairness}

\newcommand{\BibTeX}{\rm B\kern-.05em{\sc i\kern-.025em b}\kern-.08em\TeX}

\newtheorem{definition}{Definition}
\newtheorem{remark}{Remark}
\newtheorem{theorem}{Theorem}

\newtheorem{claim}{Claim}

\newcommand{\rftm}{\texttt{rTFM}}
\newcommand{\stfm}{\texttt{STFM}}

\begin{document}


\pagestyle{fancy}
\fancyhead{}



\maketitle 

\begin{textblock}{15}(0.45,1)
\centering
\noindent\small In the Proceedings of the 23\textsuperscript{rd} International Conference on Autonomous Agents and Multiagent Systems (AAMAS), 2024, as an Extended Abstract.
\end{textblock}

\section{Introduction}
\emph{Transaction Fee Mechanism} (TFM) design, introduced in the seminal work by \citet{roughgarden21}, considers the allocation problem of adding \textit{transactions} to a \textit{block} in blockchains such as Bitcoin~\cite{nakamoto2008bitcoin} and Ethereum~\cite{buterin2014next}. More concretely, the \emph{miner} of the block adds transactions to its block from the pool of outstanding transactions (aka ``mempool"). Transaction creators (henceforth \emph{users}) optionally send a \emph{transaction fee} as a commission to the miners to incentivize them to add their transactions. 

\paragraph{TFM: Framework.} The miner-user \emph{strategic} interaction in a TFM is analogous to an auction setting. Indeed, Bitcoin implements a ``first-price" auction with a miner maximizing its revenue by greedily adding transactions to its block from the mempool. A user's transaction fee captures its \emph{valuation} for its transaction's inclusion. From~\cite{roughgarden21}, TFMs comprise (i) \emph{allocation rule}, adding transactions from the mempool to a block, (ii) \emph{payment rule}, for the payment to the miner, and (iii) \emph{burning rule}\footnote{Burning refers to removing tokens from the cryptocurrency's supply forever. E.g., by transferring them to unspendable addresses that can only receive tokens, thus making the tokens inaccessible.}. Unlike classic auction settings, in TFMs, the miners have \underline{complete} control over the transactions they add. Consequently, \citet{roughgarden21} introduces \emph{miner incentive compatibility} (MIC) in addition to the standard \emph{user incentive compatibility} (UIC). MIC states that the proposed TFM must incentivize miners to follow the intended allocation rule truthfully. UIC ensures that users offer their transaction's valuation as a transaction fee. Next, we have \emph{off-chain collusion proofness} (OCAP) to curb miner-user off-chain collusion. \citet{roughgarden21} studies popular TFMs like first-price, second-price, and Ethereum's new dynamic posted-price mechanism, namely EIP-1559~\cite{buterin2019eip}, in terms of the properties they satisfy. Subsequent works \cite{parkes21,chung2021foundations} enrich the TFM literature by proposing a dynamic posted-price mechanism and providing significant foundational results, respectively.

\paragraph{TFM: Challenges with Incentives.} To satisfy UIC, MIC, and OCAP, TFMs introduce payment and burning rules based on transaction fees. However, we believe that (and as originally intended in Bitcoin~\cite{historic}) TFMs must also support including transactions with zero fees. In practice, the fees are also higher than recommended~\cite{messias2020blockchain}. Supporting zero-fee transactions will also benefit the adoption of currencies like Bitcoin and Ethereum. \textbf{First}, commission-based digital payment networks (e.g., VISA/MasterCard) are losing ground to commission-less networks (e.g., UPI)~\cite{visa}. Commission-less payment networks admit $\approx 7.5$ times \emph{higher} transaction volume compared to their commission-based counterparts~(\url{rbi.org.in}). \textbf{Second}, networks such as VISA/MasterCard charge the merchant a constant fraction of the transaction amount. This charge is \emph{unlike} Bitcoin/Ethereum, whose transaction fees are independent of the transaction amount and paid by the user. For micropayments (e.g., paying for your morning coffee), these fees are unreasonable~\cite{micropayments}. 

\subsection*{Our Approach and Contributions}
\noindent\textbf{Fairness Notions.} We introduce (i) \emph{Zero-fee Transaction Inclusion} (ZTi) and (ii) \emph{Monotonicity} (Section~\ref{sec::fairNot}). A TFM satisfies ZTi if it ensures that zero-fee transactions have a non-zero probability of getting included in the block.\footnote{We assume that miners/users are myopic~\cite{roughgarden21,parkes21,chung2021foundations}, i.e., they only consider their utility from the next block. Thus, ZTi deals with a transaction's probability of inclusion for the next block and \emph{not} ``eventual" confirmation.} However, guaranteeing ZTi must still ensure that the probability of a transaction's inclusion increases with an increase in its fee. E.g., randomly including transactions trivially ensures ZTi but may be unfair for a company that desires swift confirmation to meet the scheduled launch or if the transaction fixes a critical bug. To capture this, we introduce Monotonicity, which states that a TFM must ensure that transactions with a higher transaction fee have a greater probability of getting included in the block. Such a notion allows for \textit{priority-based} transaction confirmation. Our two fairness notions combined imply that \textit{every} transaction in the mempool has a non-zero probability of getting included in the block!

Given the impossibility of satisfying UIC, MIC, and OCAP simultaneously~\cite{chung2021foundations}, we say a TFM is \emph{fair} if it meets the above two notions, UIC and MIC. That is, fairness in TFMs w.r.t. the transaction creators (or users). Intuitively, as TFM design generally focuses on maximizing the miner's utility, it fails to satisfy ZTi. Moreover, we show that existing TFMs either do not satisfy our fairness notions or do so at a high cost to the miner's utility (Section~\ref{subsec::imp}). As such, we introduce Randomized TFM (\rftm), a TFM that satisfies our fairness notions, and study its incentive properties.

\paragraph{Randomized TFM} (\rftm). We propose \rftm (Section~\ref{sec::rtfm}), a TFM that satisfies our fairness notions while guaranteeing MIC (for an appropriate payment rule). In \rftm, we introduce a novel allocation rule that requires the miner to create two sets of transactions. In the first set, the miner optimally selects the transactions to add to its block (i.e., exactly like it currently does in Bitcoin). In the second set, the miner \emph{uniformly} adds transactions from the mempool to its block but crucially receives \emph{no} fee for these transactions. That is, the miner has no incentive to deviate from the uniform allocation in this set. The miner broadcasts both these sets, and we show that the blockchain network can randomly confirm one of the two sets through a \emph{trusted} coin-flip mechanism (Section~\ref{ssec::rtfm}). Intuitively, such an allocation gives a non-zero probability of inclusion for zero-fee transactions due to the uniform sampling in the second set. As the miner has no control over the confirmed set, \rftm\ satisfies MIC for an appropriate payment rule, e.g., Bitcoin's first-price auction (Section~\ref{subsec::rtfm_faitnot}).

\section{Related Work}\label{sec::app_rw}
We now place our work concerning the existing literature for (i) TFM design and (ii) fairness in the context of blockchain.

\smallskip
\noindent\textbf{Transaction Fee Mechanism (TFM) Design.} \citet{roughgarden21} presents the seminal work that describes the ``inclusion of transactions in a block" in the language of mechanism design. The author shows that EIP-1559 satisfies UIC and MIC and is OCAP (under some constraints on the base fee). \citet{parkes21} present a novel dynamic posted-price TFM with an equilibrium characterization of the posted-price. Most recently, \citet{chung2021foundations} provide several foundational results for TFM design based on underlying incentives and allocation rules. While the works \cite{roughgarden21,parkes21,zhao2022bayesian,chung2021foundations} are complementary, they do not focus on transaction fairness in TFMs.

Parallely, works also exist that empirically analyze TFMs to optimize transaction fees~\cite{laurent2022transactions,tedeschi2022optimizing}. \citet{tedeschi2022optimizing} suggest a Deep Neural Network-based approach to predict miners' behavior in terms of including transactions in their blocks. The authors show that their approach reduces transaction fees and improves the confirmation time. 

\paragraph{Fairness in Blockchain.} Fairness is studied in various contexts, including network latency~\cite{jain21,mao2022less}, transaction ordering~\cite{gervais2014privacy,asayag2018fair,sokolik2020age,orda2021enforcing,kelkar2020order,kursawe2020wendy} and price of transaction consumption~\cite{basu2019towards,bitcoinf}. 

Fairness in transaction order focuses on the latency in transaction confirmation. E.g., miners may discriminate among specific transaction creators or only include transactions of the creators they know prior. This line of work~\cite{gervais2014privacy,asayag2018fair,sokolik2020age,orda2021enforcing,kelkar2020order,kursawe2020wendy} does not model game-theoretic interactions and focuses on verifiable methods of ensuring ``fairness" using cryptographic primitives. Moreover, there is no provision for the inclusion of zero-fee transactions. E.g., \citet{sokolik2020age} present a fair approach that prioritizes transactions with significant waiting time. \citet{orda2021enforcing} provide techniques that enforce that transactions are allocated randomly to each block.

BitcoinF's~\cite{bitcoinf} allocation rule splits the block with dedicated sections for standard transactions and low-fee transactions. The authors argue that this allows miners to maximize their utility (through the standard section) while also processing low-fee transactions. With a strong assumption that transaction influx equals the cryptocurrency's throughput, they empirically argue that BitcoinF provides a lower consumption price. Also, they do not provide any theoretical guarantees for strategyproofness or fairness.

\section{Preliminaries}\label{sec::preli}
We now summarize (i) the TFM and user model,  (ii) relevant game-theoretic definitions, (iii) existing TFMs, and (iv) required blockchain preliminaries.

\subsection{TFM Model}
TFM design for public blockchains such as Bitcoin~\cite{nakamoto2008bitcoin} and Ethereum~\cite{buterin2014next} considers the following model. The blockchain's public ledger maintains the \emph{state} and orders the sequence of \emph{transactions} $t_1, t_2, \ldots, t_n, n\in\mathbb{N}_{\geq 1}$ that update the state. Let $s_i\in\mathbb{R}_{>0}$ be the size\footnote{E.g., Ethereum transactions may be token transfers (smaller size) or sophisticated smart contract calls (larger size).} of a transaction $t_i$. Each user $i$ broadcasts its transaction $t_i$ with a bid (per unit size) $b_i\in\mathbb{R}_{\geq 0}$. That is, the total bid is $s_i \cdot b_i$. The bid represents the amount user $i$ is willing to pay for $t_i$, given its (per unit size) \emph{private} valuation $\theta_i \in\mathbb{R}_{\geq 0}$. For security and practical reasons, each block has a \emph{finite} capacity (denoted by $C\in\mathbb{R}_{>0}$). Miners create blocks, maintain a \emph{mempool} of outstanding transactions $(M:=\{t_1,\ldots,t_n\})$, and add a subset of these transactions to their blocks. Generally, the set of outstanding transactions is larger than the block size.

\paragraph{Transaction Fee Mechanism (TFM).} Consider $\mathcal{H}=B_1,\ldots,B_{k-1}$ as the sequence of blocks denoting the on-chain history, current block $B_k$ and mempool $M$. Designing a TFM involves defining (i) an \emph{allocation} rule, which decides the transactions that get added to $B_k$, (ii) a \emph{payment} rule describing the fraction of each transaction's bid that gets paid to the miner, and (iii) a \emph{burning} rule, that is, the fraction of the amount that is removed from the supply, forever. An idiosyncrasy of blockchain involves \emph{randomization} in transaction allocation. More concretely, with a ``deterministic" TFM, we imply that a miner can include transactions in its block using any deterministic function. Whereas a ``randomized" TFM implies that the miner selects the transactions to include through a random function\footnote{TFMs may also use trusted on-chain randomness for transaction inclusion~\cite{chung2021foundations}.}. To the TFM definition proposed in \cite{roughgarden21}, we explicitly add the provision of TFMs being randomized. 
\begin{definition}[Transaction Fee Mechanism (TFM)]\normalfont \label{def:tfm} For a given on-chain history $\mathcal{H}$, the mempool $M$ and the current block $B_k$ with size $C$, a TFM is the tuple $\mathcal{T}^{TFM}=(\mathbf{x},\mathbf{p},\mathbf{q},\tau)$ in which,
    \begin{enumerate}[leftmargin=*]
        \item $\mathbf{x}$ is a feasible block allocation rule, i.e., $\sum_{t\in M}s_t\cdot x_t(\mathcal{H},M)\leq C$ where $x_t(\cdot)\in\{0,1\}$,~$\forall t\in M$.
        \item $\mathbf{p}$ is the payment rule with the payment for each transaction $t\in B_k$ denoted by $p_t(\mathcal{H},B_k)\geq 0$.
        \item $\mathbf{q}$ is the burning rule with the amount of burned coins for each transaction $t\in B_k$ denoted by $q_t(\mathcal{H},B_k)\geq 0$.
        \item $\tau\in\{\tau_D,\tau_R\}$ is the mechanism's type -- either deterministic ($\tau_D$) or randomized ($\tau_R$).
    \end{enumerate}
\end{definition}


\subsection{User Model and Incentive Properties}
We now define the relevant incentive properties introduced in \cite{roughgarden21} for a TFM. We assume that the miners and bidding users are myopic~\cite{roughgarden21,parkes21,zhao2022bayesian,chung2021foundations} --  they are only concerned with their utility from the next block. For each user $i$, we have its (per unit size) valuation $\theta_i$, its bid $b_i$, and transaction size $s_i$. Let the vector $\mathbf{b}$ comprise all bids with $\mathbf{b}_{-i}$ representing all bids without user $i$. Given $\mathcal{T}^{TFM}=(\mathbf{x},\mathbf{p},\mathbf{q},\tau)$ with $\mathcal{H},M,\mbox{~and~}B_k$, an user $i$'s \textit{quasi-linear} utility $u_i$ is,
    \begin{equation}\label{eqn::util}
    u_i(\mathbf{b}) :=
        \begin{cases}
         \left(\theta_i-p_i(\cdot)-q_i(\cdot)\right) s_i \mbox{~\textbf{if}~} x_i = 1 \\
         0 \qquad \mbox{\textbf{otherwise}.}
        \end{cases}
    \end{equation}

\paragraph{User Incentive Compatibility (UIC).} A strategic user $i$ will select $b_i$ such that it maximizes its utility defined in Eq.~\ref{eqn::util}. As such, we now define UIC for a TFM.

\begin{definition}[UIC~\cite{roughgarden21}]\label{def::dsic}
A TFM $\mathcal{T}^{TFM}=(\mathbf{x},\mathbf{p},\mathbf{q},\tau)$ with $\mathcal{H}$, $M$, and $B_k$ is UIC if -- assuming the miner follows the allocation rule $\mathbf{x}$ -- bidding $\theta_i$ for each user $i$ maximizes $u_i$ (Eq.~\ref{eqn::util}), irrespective of the remaining bids. That is, $\forall i,$ $u_i(b_i^\star=\theta_i,\mathbf{b}_{-i})\geq u_i(b_i,\mathbf{b}_{-i}),$ $\forall b_i$ and  $\forall \mathbf{b}_{-i}$.
\end{definition}

Informally, UIC states that it is the best response for a user to submit its valuation as its transaction fee. 

\paragraph{Myopic Miner Incentive Compatibility (MIC).} In TFMs, the miner of block $B_k$ has complete control over the set of transactions to add to $B_k$ (i.e., implement an alternate allocation rule over the intended one). To deviate from the intended rule $\mathbf{x}$, a miner typically adds ``fake" transactions to the mempool. For the set of fake transactions $F$ (i.e., $F\subset M$) and for any $\mathcal{T}^{TFM}=(\mathbf{x},\mathbf{p},\mathbf{q},\tau)$ with $\mathcal{H}$, $M$, and $B_k$ we can write miner's utility $u_\textsf{M}$ as follows~\cite{roughgarden21}. We have $B_k=\{t\in M ~|~ x_t=1\}.$
\begin{equation}\label{eqn::min_util}
    u_\textsf{M}(B_k,F):=\sum_{t\in B_k\cap M\setminus F}
    s_t\cdot p_t(\cdot)-\sum_{t\in B_k\cap F} s_t\cdot q_t(\cdot). 
\end{equation}

The first term represents the miner's revenue, and the second term represents the fee burned from the miner's fake transactions. To maximize its utility, the miner performs the following optimization.
\begin{align}\label{eqn::miner_opt}
\begin{cases}
    \max_{\mathbf{x^\prime}} & \sum_{t\in B_k\cap M\setminus F}
    x_t\cdot s_t\cdot p_t(\cdot)-\sum_{t\in B_k\cap F} x_t\cdot s_t\cdot q_t(\cdot)\\
    \mbox{s.t.~} & \sum_{t\in M}s_t\cdot x_t\leq C\mbox{~and~} x_t(\mathcal{H},M)\in\{0,1\}, \forall t
\end{cases}
\end{align}

Given the possibility of a miner's strategic deviation, \citet{roughgarden21} introduces MIC.

\begin{definition}[MIC~\cite{roughgarden21}]\label{def::mic}
A TFM $\mathcal{T}^{TFM}=(\mathbf{x},\mathbf{p},\mathbf{q},\tau)$ with $\mathcal{H}$, $M$, and $B_k$ is MIC, if a miner maximizes $u_\textsf{M}$ (Eq.~\ref{eqn::miner_opt}) by not creating any fake transactions, $F=\emptyset$ and following the rule $\mathbf{x}$.
\end{definition}

 Let \texttt{OPT} denote the miner's optimal utility from Eq.~\ref{eqn::miner_opt} (i.e., with $p_t=b_t\mbox{~and~}q_t=0,~\forall t\in B_k$). Note that computing the optimal feasible set, say $\mathbf{x}^\star$, in Eq.~\ref{eqn::miner_opt} is NP-Hard since it reduces to KNAPSACK auctions~\cite{knapsack}. Miners may instead adopt a greedy-based approach~\cite{roughgarden21}.
 
\paragraph{Off-chain Collusion Proof (OCAP).} Another desirable property in TFM is {OCAP}, which deals with the off-chain collusion of the miner and a set of $c\in\mathbb{N}_{\geq 1}$ users. A TFM is $c$-OCAP if any coalition between the miner and set of users with cardinality $c$ Pareto improves the intended allocation $\mathbf{x}$. As stated earlier, \citet{chung2021foundations} prove the impossibility of simultaneously satisfying UIC and 1-OCAP; thus, we focus only on MIC and UIC in this work.

\subsection{Popular TFMs and Their Properties\label{subsec:pop-tfm}}
We now summarize some popular TFMs in literature. 

\paragraph{First-price (FPA) TFM.} Bitcoin employs a first-price TFM which can be expressed in the language of Definition~\ref{def:tfm} with $\mathcal{T}^{\texttt{FPA}}=(\mathbf{x}^{\texttt{FPA}},\mathbf{p}^{\texttt{FPA}},\mathbf{q}^{\texttt{FPA}},\tau^{\texttt{FPA}})$. Here, $\mathbf{x}^{\texttt{FPA}}$ follows Eq.~\ref{eqn::miner_opt}. For each $t_i\in B_k$ we have, $p_i^{\texttt{FPA}}=b_i$, $q_i^{\texttt{FPA}}=0$ and $\tau^{\texttt{FPA}}=\tau_D$. FPA does not satisfy UIC but satisfies MIC~\cite{roughgarden21}.

\paragraph{Second-price (SPA) TFM.} We denote the second-price TFM with $\mathcal{T}^{\texttt{SPA}}=(\mathbf{x}^{\texttt{SPA}},\mathbf{p}^{\texttt{SPA}},\mathbf{q}^{\texttt{SPA}},\tau^{\texttt{SPA}})$. Here, $\mathbf{x}^{\texttt{SPA}}$ follows Eq.~\ref{eqn::miner_opt}. Assuming $\bar{b}$ as the lowest winning bid, for each $t_i\in B_k$, we have\footnote{Generally, SPAs require users to pay the highest losing bid. As payments cannot depend on transactions not part of a block, \citet{roughgarden21} suggests using the lowest winning bid as a proxy.}, $p_i^{\texttt{SPA}}=\bar{b}$, $q_i^{\texttt{SPA}}=0$ and $\tau^{\texttt{SPA}}=\tau_D$. SPA approximately satisfies UIC but does not satisfy MIC~\cite{roughgarden21}.

\smallskip
\noindent\textbf{EIP-1559~\cite{buterin2019eip}.} Denoted with $\mathcal{T}^{1559}=$ $(\mathbf{x}^{1559},\mathbf{p}^{1559},\mathbf{q}^{1559},\tau^{1559})$, in EIP-1559, for each $t_i\in B_k$, we have $p^{1559}_{i}(\mathcal{H},B_k)=b_{i}-\lambda$ where $\lambda$ is the (dynamic) base fee\footnote{$\lambda$ is dynamic and depends on the network congestion. If the block size $>C$, the congestion is higher, and $\lambda$ is incremented by 12.5\%. If the block size is $\leq C$, $\lambda$ is decremented by 12.5\%~\cite{roughgarden21}.}, $q^{1559}_i=\lambda$ and $\tau^{1559}=\tau_D$. The miner maximizes its utility such that $\mathbf{x}^{1559}$ follows Eq.~\ref{eqn::miner_opt}.

EIP-1559 satisfies UIC \emph{only} if $\lambda$ is \emph{not} ``excessively low" \cite[Def. 5.6]{TFM_full}. The base fee $\lambda$ is excessively low if $\lambda$ is large enough so that the number of transactions with a valuation greater than $\lambda$ does not exceed the block size. EIP-1559 also satisfies MIC. 

\smallskip
\noindent\textbf{BitcoinF~\cite{bitcoinf}.} We denote BitcoinF as $\mathcal{T}^{B}=(\mathbf{x}^{B},\mathbf{p}^{B},\mathbf{q}^{B},\tau^{B})$. Each user $i$ creates \emph{two} transactions offering a public constant fee $\delta\in\mathbb{R}_{>0}$ and $\delta+\hat{b_i},\hat{b_i}\in\mathbb{R}_{>0}$ as fees. If one gets added, the other is nullified. The allocation rule $\mathbf{x}^{B}$ splits the block into $\alpha\in (0,1]$ and $1-\alpha$ fractions. The miner must first fill the $1-\alpha$ section through FIFO collecting transactions with $\delta$, after which it can greedily fill the $\alpha$ section. 
Let $C_{\alpha}$ and $C_{1-\alpha}$ denote the capacity of the $\alpha$ and $1-\alpha$ sections, i.e., $C=C_\alpha + C_{1-\alpha}$. 
For each $t_i$ in the $\alpha$ section, we have $p_i^B=\hat{b}_i+\delta$ and $q_i^B=0$. Likewise, for each $i$ in the $1-\alpha$ section, we have $p_i^B=\delta$ and $q_i^B=0$. Lastly, $\tau^B=\tau_D$. BitcoinF's optimization is as follows.

\begin{align}\label{eqn::BF_opt}
\begin{rcases}
    \max_{\mathbf{x}^B} & \sum_{i\in M} x_i^B\cdot p_i^B(\mathcal{H},B_k)\cdot s_i\\
    \mbox{s.t.~} & \sum_{t\in M, b_t\not=\delta}s_t\cdot x_t^B(\mathcal{H},M) \leq C_{\alpha} \\
    & \sum_{t\in M, b_t=\delta}s_t\cdot x_t^B(\mathcal{H},M) = C_{1-\alpha} \mbox{~and~}  \\
  & x_t^B(\mathcal{H},M)\in\{0,1\}, \forall t\in M.
\end{rcases}
\end{align}

As a warm-up result, we show that strategic miners in $\mathcal{T}^{B}$ may deviate, i.e., miners may include fake transactions in the $1-\alpha$ section of the block to increase their utility from the $\alpha$ section. Remark~\ref{rem::bf} captures this result. For the proof, in Appendix~\ref{remark1}, we construct an example showing that a miner can add fake transactions to increase utility. 

\begin{remark}\label{rem::bf}
BitcoinF $(\mathcal{T}^{B})$ does not satisfy MIC. 
\end{remark}

Section \ref{sec::rtfm} presents a novel TFM -- namely, \rftm -- that leverages specific blockchain and cryptographic fundamentals, as outlined next.

 \subsection{Blockchain and Cryptographic Preliminaries}

\paragraph{Hash Functions.} Given a security parameter $\lambda\in\mathbb{N}_{\geq 1}$, cryptographic hash functions are one-way functions defined as $\textsc{Hash}:\{0,1\}^* \rightarrow \{0,1\}^\lambda$. A hash function is (i) \textit{collision-resistant} if the probability of any two distinct inputs $x,y$ map to the same output with negligible probability, i.e., $\Pr[\textsc{Hash}(x)=\textsc{Hash}(y)|x\not = y] \leq \textsf{negl}(\lambda)$ and (ii) \textit{pre-image resistant} if the probability of inverting $\textsc{Hash}(x)$ is less than $\textsf{negl}(\lambda)$. Here, $\textsf{negl}(\lambda)$ denotes a negligible function in $\lambda$. E.g., SHA-256~\cite{gilbert2003SHA}.



\paragraph{Merkle Tree (MT)~\cite{merkle1987digital}.} These are complete binary trees where every parent node is a hash of its children. In blockchains like Bitcoin, each block comprises an MT such that the parents are hashes of transactions that are included in the block. More concretely, the value of a parent node $a$ is the hash of the concatenation of its two children nodes $b,c$, i.e., $a = \textsc{Hash}(b||c)$. The Merkle root $\textsf{root}$ is the hash value of the root node of MT.

\paragraph{Proof-of-Work (PoW)~\cite{nakamoto2008bitcoin}.} In blockchains like Bitcoin~\cite{nakamoto2008bitcoin}, PoW is a protocol to propose new blocks. Here, miners use the blockchain's history $\mathcal{H}$ (comprising previously mined blocks, say up till $B_{k-1}$) and $\textsf{root}$ of the set of transactions to be included in their block, $B_k$. The block header of $B_{k}$ is made up of the hash of the parent block $B_{k-1}$, $\textsf{root}$, and a randomly generated nonce. The block is considered mined if the miner finds a nonce such that the hash value of the block $h = \textsc{Hash}(B_{k})$ is lesser than \emph{target difficulty} ($TD$) as decided by the system, i.e., $h < TD$.

\paragraph{On-chain Trusted Randomness.} \citet{micali1999verifiable} introduce \emph{verifiable random functions}, which take inputs and generate pseudorandom outputs that can be publicly verified. In the blockchain context, this often implies functions whose randomness depends on the information available to the blockchain (aka verifiable or trusted on-chain randomness). E.g., \citet{chung2021foundations} propose a randomized second-price TFM that uses such randomness to confirm transactions added to its block by the miner. 


\section{Fairness in TFMs}\label{sec::fairNot}

This section (i) presents our novel fairness notions, (ii) proves the impossibility of simultaneously maximizing the miner's utility and ZTi, (iii) studies the fairness guarantees of BitcoinF when $\delta=0,$ and (iv) discusses Softmax TFM (\stfm).

\subsection{Fairness Notions}\label{ssec::fairness-notions}
We propose the following fairness notions to tackle the challenges due to transaction fees in TFMs. 

\smallskip
\noindent\circled{1}~\textbf{Zero-fee Transaction Inclusion (ZTi).} In Bitcoin, a TFM requires a user to pay transaction fees, even for micropayments. Furthermore, there is an unbounded waiting time for transactions with marginal fees in Bitcoin~\cite{bitcoinf}. As such, we introduce \emph{Zero-fee Transaction Inclusion} (ZTi) as a critical fairness notion for a TFM to satisfy. That is, our first fairness notion ensures that a transaction with zero fees must have a non-zero probability of getting included in the block.

\begin{definition}[Zero-fee Transaction Inclusion (ZTi)]\label{def::zero}
 A TFM $\mathcal{T}^{TFM}$ satisfies ZTi if the probability with which a transaction $t$ with transaction fee $b_t=0$ gets included in a block $B_k$ is strictly non-zero, i.e., $\Pr(t\in B_k) > 0$.
\end{definition}

As the users and miners are myopic, ZTi only considers a transaction's probability of being included in the next block.

\smallskip
\noindent\circled{2}~\textbf{Monotonicity.} This notion focuses on the probability of the inclusion of a bidding user's transaction being proportional to the transaction fee. Naturally, a user would expect a higher probability of its transaction being included if it increases the transaction's fee. Such a scenario is also desirable in practice, e.g., startups/applications may want faster transaction acceptance to meet launch dates, deployment targets, or critical bug fixes. 

\begin{definition}[Monotonicity]\label{def::prop}
 a TFM $\mathcal{T}^{TFM}$ satisfies Monotonicity if the probability with which a transaction $t$ gets accepted in a block $B_k$ increases with an increase in its transaction fee $b_t$, given the remaining bids $\mathbf{b}_{-t}$ are fixed. That is, $\Pr(t\in B_k ~|~ \mathbf{b}_{-t}, b_t+\epsilon ) > \Pr(t\in B_k ~|~\mathbf{b}_{-t}, b_t )$ for any $\epsilon > 0$ and fixed $\mathbf{b}_{-t}$.
\end{definition}

We remark that most existing TFMs satisfy monotonicity. However, designing TFMs that satisfy monotonicity and ZTi simultaneously is non-trivial. Trivially, a TFM satisfying both our fairness notions ensures that each transaction has a non-zero probability of getting accepted!

\subsection{Impossibility of Simultaneously Maximizing Miner Utility and Satisfying ZTi\label{subsec::imp}}

 Before presenting the main impossibility, we first analyze the fairness guarantees for EIP-1559~\cite{buterin2019eip}.

\begin{remark}\label{remark::notion}\normalfont
EIP-1559 satisfies (i) Monotonicity but does not satisfy (ii) ZTi. As each transaction must at least pay the base fee, no honest/strategic miner will include zero-fee transactions to preserve the validity of their blocks, i.e., if $b_t=0\implies \Pr(t\in B)=0$. EIP-1559 satisfies monotonicity since increasing the payment $b_t-\lambda$ will increase the chance of the transaction being part of the optimal set in Eq.~\ref{eqn::miner_opt}.
\end{remark}

Theorem~\ref{thm::imp} adds to Remark~\ref{remark::notion} by showing that any TFM that allows a strategic miner complete control over which transactions to add cannot satisfy ZTi for any non-trivial payment rule. A \emph{trivial payment rule} is $p_t=0,~\forall t\in B_k$. For the proof, in Appendix~\ref{app:thm-imp}, we provide a counterexample s.t. $\forall t\in M,~b_t=0\implies\Pr(t\in B_k=0)$. 

\begin{theorem}\label{thm::imp}
No $\mathcal{T}^{TFM}$ with a non-trivial payment rule, which provides a strategic miner complete control over the transactions to add to its block, satisfies Zero-fee Transaction Inclusion (ZTi).
\end{theorem}

\subsection{BitcoinZF: BitcoinF with Zero Fees\label{subsec::zf}}
We tweak the block allocation rule in BitcoinF~\cite{bitcoinf} to introduce a provision for transactions with zero fees. We set $\delta=0$ so that the miner \emph{randomly} adds zero-fee transactions to fill the $1-\alpha$ section, followed by \textit{greedily} adding transactions with bid $b$ to the $\alpha$ section. The formal optimization can be derived by fixing $\delta=0$ in Eq.~\ref{eqn::BF_opt}.

\begin{align}\label{eqn::new_BF_opt}
\begin{rcases}
    \max_{\mathbf{x}^{BZ}} & \sum_{i\in M} x_i^{BZ}\cdot p_i^{BZ}(\mathcal{H},B_k)\cdot s_i\\
    \mbox{s.t.~} & \sum_{t\in M, b_t\not=0}s_t\cdot x_t^{BZ}(\mathcal{H},M) \leq C_{\alpha} \\
    & \sum_{t\in M, b_t=0}s_t\cdot x_t^{BZ}(\mathcal{H},M) = C_{1-\alpha} \mbox{~and~}  \\
   & x_t^{BZ}(\mathcal{H},M)\in\{0,1\}, \forall t\in M.
\end{rcases}
\end{align}

Furthermore, with base fee $\lambda$, for each $i$ in the $\alpha$ section we have $p_i^{BZ}=b_i-\lambda$ and $q_i^{BZ}=\lambda$. For each $i$ in the $1-\alpha$ section we have $p_i^{BZ}=q_i^{BZ}=0$. In summary, BitcoinZF is denoted by the tuple $\mathcal{T}^{BZ}=(\mathbf{x}^{BZ},\mathbf{p}^{BZ},\mathbf{q}^{BZ},\tau_D)$.

\paragraph{Fairness Notions.} Theorem~\ref{thm::BF-Fairness} shows that BitcoinZF satisfies the two fairness notions if each zero-fee transaction's size is less than $C_{1-\alpha}$. In other words, BitcoinZF satisfies ZTi if none of the zero-fee transactions are of significant size.

\begin{theorem}\label{thm::BF-Fairness}
BitcoinZF $(\mathcal{T}^{BZ})$ satisfies (i) Zero-fee Transaction Inclusion and (ii) Monotonicity only if $\forall~ t_i\in M$ with $b_i=0$, we have $s_i\leq C_{1-\alpha}$.
\end{theorem}

We defer Theorem~\ref{thm::BF-Fairness}'s proof to Appendix~\ref{app:thm-BF-Fairness} in the supplementary. Informally, let a user $i$ increase its $b_i$. At the same time, if the other bids remain unchanged, user $i$'s chances of being included in the ``$\alpha$" section increase, satisfying Monotonicity. Furthermore, since the miner receives no increase in utility from any transaction in the ``$1-\alpha$" section, it can uniformly include zero-fee transactions. 

\smallskip
\noindent\textbf{Cost of Fairness}~\textsf{(CoF)}. Unfortunately, there is a ``cost" to the fairness guarantees in BitcoinZF. Ensuring ZTi \emph{hurts} the miner's utility. To this end, consider the following definition.


\begin{definition}[\textsf{CoF}]\label{def::cof}
We define \textsf{(CoF)} of $\mathcal{T}^{TFM}=(\mathbf{x},\mathbf{p},\mathbf{q},\tau)$ as \textsf{CoF}$_{TFM}=\max_{\mathbf{b}\neq 0}\frac{\texttt{OPT}}{u_\textsf{M}^{TFM}}$. Here, $u_\textsf{M}^{TFM}$ is the miner's utility from the indented allocation $\mathbf{x}$ and \texttt{OPT} its utility from Eq.~\ref{eqn::miner_opt} with $p_t=b_t\mbox{~and~}q_t=0,~\forall t\in B_k$.
\end{definition}

Trivially, \emph{lesser} the \textsf{CoF}, \emph{greater} the miner's utility from following $\mathcal{T}^{TFM}$.
Claim~\ref{claim::BF_bound} presents the \textsf{CoF} for BitcoinZF for the specific case when for every $t_i,t_j\in M$ s.t. $i\not=j$, we have $s_i=s_j$. That is, all transactions are of the same size. The proof follows from algebraic manipulations; refer to Appendix~\ref{app:claim-BF-bound} of the supplementary.

\begin{claim}\label{claim::BF_bound}
For every $t_i,t_j\in M$ s.t. $i\not=j$, if we have $s_i=s_j$, then \textsf{CoF}$_{BZ}=\frac{\texttt{OPT}}{u_\textsf{M}^{BZ}}=1/\alpha$ where $\alpha\in (0,1]$.
\end{claim}

\noindent\textbf{Challenges with BitcoinZF.} Despite satisfying our fairness notions, BitcoinZF has the following challenges. First, Claim~\ref{claim::BF_bound} only holds when each transaction's size is equal. With different transaction sizes,  $\frac{\texttt{OPT}}{u_\textsf{M}^{BZ}}$  can be arbitrarily bad. E.g., if the size of the transaction with the highest bid in $M$ is greater than $C_\alpha$, ${\texttt{OPT}}/{u_\textsf{M}^{BZ}}\to \infty$. Second, when $1-\alpha$ is small, zero-fee transactions of sufficient size will deterministicly never get included in the block.
Formally, if $\exists~ t_i\in M$ s.t. $b_i=0$ and $s_i> C_{1-\alpha}$, we have $\Pr(t_i\in B_k)= 0$.

To this end, we next propose a novel TFM with randomized allocation using the softmax with temperature function.  

\section{\stfm: First Approach to Achieve Fairness Through Randomization\label{subsec::soft_stfm}}
We now introduce Softmax TFM (STFM), which comprises an intuitive, randomized allocation rule that guarantees ZTi and Monotonicity. To begin with, it's important to note that a straightforward allocation rule that uniformly selects transactions from $M$ will trivially satisfy both our fairness notions.

\begin{remark}\normalfont\label{rem3}
Consider a TFM with an allocation rule that uniformly samples transactions, i.e., $\forall t_i\in M$ $\Pr(t_i\in B_k)=1/n$. Trivially, such a TFM (i) satisfies Zero-fee Transaction Inclusion but (ii) does not satisfy Monotonicity since as $b_i$ increases, $\Pr(t_i\in B_k)$ remains the same. 
\end{remark}

We next (i) introduce Softmax TFM (STFM) and (ii) discuss its fairness and incentive guarantees.

\subsection{Softmax TFM}
For a given on-chain history $\mathcal{H}$, the mempool $M$ and the current block $B_k$, STFM can be expressed in the TFM language as $\mathcal{T}^{STFM}=(\mathbf{x}^{STFM},\mathbf{p}^{STFM},\mathbf{q}^{STFM},\tau_R)$. We begin by defining the allocation rule $\mathbf{x}^{STFM}$.


\paragraph{STFM Allocation.} Unlike deterministic TFMs like FPA and EIP-1559, STFM is a randomized allocation rule. The miner does not compute the optimal allocation set as in Eq.~\ref{eqn::miner_opt} but instead \textit{samples} a feasible set of transactions. These transactions are sampled through a distribution generated by applying the softmax with temperature function to the set of the outstanding transactions in $M$. The softmax function with the temperature parameter $\gamma\in\mathbb{R}^+$ and for any real-valued vector $\mathbf{z}=(z_1,\ldots,z_n)$ is defined $\forall i$ as follows.
\begin{equation}\label{eqn::softmax}
    \Gamma(\mathbf{z})_i = \frac{\exp(z_i/\gamma)}{\sum_{i^\prime \in \mathbf{z}}\exp(z_{i^\prime}/\gamma)}.
\end{equation}

Algorithm~\ref{algo::sftm} presents the procedure with which the miner randomly samples a feasible set of transactions in STFM. With this, we can define $\mathbf{x}^{STFM}$ as follows.

\begin{definition}[STFM Allocation Rule]\label{def::sftm_alloc}
Given $\mathcal{H}$, $M$ and $B_k$, let $\mathbf{x}^{STFM}$ denote a feasible allocation rule with $\Pr(t\in B_k),~\forall t\in M$ generated from the Softmax distribution (refer Algorithm~\ref{algo::sftm}). Formally, given the set of transactions sampled, $\mathcal{X}_k\leftarrow\textsc{STFMAllocation}(C,M,\mathcal{H})$, we have $\mathbf{x}^{STFM}=[x_t^{STFM}]$ s.t.
    \begin{equation}
    x_t^{STFM} =
        \begin{cases}
         1 \qquad \mbox{if} \quad t  \in \mathcal{X}_k,\\
         0 \qquad \mbox{otherwise.}
        \end{cases}
    \end{equation}
\end{definition}
    
    


\paragraph{STFM Payment and Burning Rules.} The allocation rule $\mathbf{x}^{STFM}$ can be coupled with any payment $(\mathbf{p}^{STFM})$ and burning $(\mathbf{q}^{STFM})$ rules to define $\mathcal{T}^{STFM}$. E.g., similar to FPA, we can create $\mathcal{T}^{STFM}$ such that each bidding agent $i$ whose $t_i\in B_k$ pays $p_i^{STFM}=b_i$ and $p_i^{STFM}=0$ otherwise. Furthermore, $q_i^{STFM}=0,~\forall i$.

\subsection{STFM: Fairness Properties} The choice of the payment and burning rules impact the strategyproofness, w.r.t. both the agent and the miner, of the resulting STFM mechanism. However, Theorem~\ref{thm::sftm_fair} proves that the STFM allocation from Definition~\ref{def::sftm_alloc} is sufficient to satisfy both our fairness notions. 

\begin{theorem}\label{thm::sftm_fair}
$\mathcal{T}^{STFM}$ with $\gamma\in (0,\infty)$ satisfies (i) Zero-fee Transaction Inclusion and (ii) Monotonicity.
\end{theorem}

\begin{algorithm}[!t] 
\caption{\label{algo::sftm} Softmax TFM (STFM) Allocation}
\begin{algorithmic}[1]
\Statex\textbf{Input:} Block Size $C$, Mempool $M$, History $\mathcal{H}$, Temperature $\gamma$
\Statex\textbf{Output:} Set of allocated transactions in $B_k$, i.e., $\mathcal{X}_k$
\Procedure{STFMAllocation}{$C,M,\mathcal{H}$}
\State $S = 0, \mathcal{X}_k=\emptyset$
\State $\Gamma_k=\left[\frac{\exp(b_t/\gamma)}{\sum_{t^\prime \in M}\exp(b_{t^\prime}/\gamma)}\right]_{\forall t\in M}$ \Comment{\textcolor{blue}{Generate the Softmax distribution}}
\While{$C-S>0$}
    \State $t \sim \Gamma_k$ \Comment{\textcolor{blue}{Sample a transaction}}
    \State $S\leftarrow S + s_t$ \Comment{\textcolor{blue}{Add to the current block consumption}}
    \State $\mathcal{X}_k\leftarrow \mathcal{X}_k+\{t\}$
    \State $\Gamma_k=\left[\frac{\exp(b_t/\gamma)}{\sum_{t^\prime \in M\setminus\mathcal{X}_k}\exp(b_{t^\prime}/\gamma)}\right]_{\forall t\in M\setminus\mathcal{X}_k}$ \Comment{\textcolor{blue}{Re-generate the Softmax distribution}}
\EndWhile
\State \Return $\mathcal{X}_k$
\EndProcedure
\end{algorithmic}
\end{algorithm}


\subsection{Softmax TFM: Incentive Properties\label{subsec::sftm-ic}} As aforementioned, the incentive properties of STFM are a function of the underlying payment and burning rules. Theorem~\ref{thm::sftm-mic-general} presents the general impossibility of MIC for STFM with any payment rule, which increases monotonically with the transaction fees. For the proof, we show that for any non-trivial payment rule, $\mathbf{x}^{STFM}$ in $\mathcal{T}^{STFM}$ is such that the miner has an incentive to deviate.

\begin{theorem}\label{thm::sftm-mic-general}
Given $\gamma\in (0,\infty)$ and with any non-trivial, monotonically increasing $\mathbf{p}^{STFM}$, i.e., for $b_i>b_j\implies p_i>p_j$ $\forall i,j \mbox{~s.t.~}i\not=j$, $\mathcal{T}^{STFM}$ does not satisfy MIC.
\end{theorem}

 We now discuss UIC and MIC guarantees for STFM with FPA and EIP-1559 payment and burning rules. 

\begin{remark}\normalfont\label{rem4}
From the perspective of the bidding agent, STFM's allocation rule does not change its behavior as the allocation rule satisfies Monotonicity. As such, any instance of STFM with FPA does not satisfy UIC, as the first-price payment rule is well-known not to be UIC. Furthermore, STFM with EIP-1559 satisfies UIC only when the underlying EIP-1559 is UIC (refer to Section~\ref{subsec:pop-tfm}). 
\end{remark}

\paragraph{STFM: Cost of Fairness.} Similar to \textsf{CoF} guarantees for BitcoinZF, we next provide an upper bound on \textsf{CoF} for STFM. We obtain the bound by selecting the worst-case distribution of bids which maximize $\mathbb{E}[{OPT}]$ and minimize $\mathbb{E}[u_m^{STFM}]$. 

\begin{theorem}\label{thm::stfm-upperlim}
For STFM with FPA, average \textsf{CoF}$_{STFM}=\frac{OPT}{\mathbb{E}_{\mathbf{x}\sim \Gamma_k}[u_m^{STFM}]}=\frac{n}{c} + 1$. Here, $n$ denotes the total transactions in $M$ and $c$ the maximum number of transactions included in $B_k$. 
\end{theorem}

\paragraph{Note.} Despite STFM satisfying ZTi and Monotonicity, Theorem~\ref{thm::sftm-mic-general} states that it is not MIC under any monotone payment rule. To this end, we next leverage the blockchain's verifiable randomness to propose \rftm\, a TFM that satisfies both our fairness notions while simultaneously guaranteeing MIC.

\subsection{Softmax TFM: Tuning $\gamma$ for Increased Miner Utility}
We observe that satisfying our fairness notions with STFM reduces a miner's utility. Naturally, each miner of a block will prefer to increase its utility. We now discuss the role of the temperature parameter $\gamma$ in improving the miner's utility while simultaneously retaining the fairness guarantees.

\paragraph{STFM at $\gamma\to\infty$.} Observe that, from \eqref{eqn::softmax}, as $\gamma$ increases, the softmax probability distributions tend toward the uniform distribution. When $\gamma\to\infty$, the distribution becomes Uniform, i.e., all transactions are included with the same probability. That is, at $\gamma\to\infty$, STFM does not satisfy Monotonicity, and the miner's utility loss is at its maximum.

\paragraph{STFM at $\gamma\to 0$.} In contrast to the previous scenario, when $\gamma\to 0$, STFM's allocation mimics the optimal allocation from \eqref{eqn::miner_opt}. That is, at $\gamma\to 0$, STFM does not satisfy ZTi, and the miner's utility loss is approximately zero.

\paragraph{An Improved Trade-off.} In Appendix~\ref{ssec::results}, we show how to derive an ideal value of $\gamma^\star$ with regards to \textsf{CoF} and number of zero-fee transactions included. More concretely, we first derive the expression of the ratio of the probability of the optimal set of transactions (from Eq.~\ref{eqn::miner_opt}) being included to the block with probability of some $\alpha\in[0,1]$ fraction of block comprising transactions with zero-fees (say $\frac{pr_\textsf{CoF}}{pr_\textsf{ZF}}$). Then, we solve for $\gamma^\star$ s.t. $\frac{pr_\textsf{CoF}}{pr_\textsf{ZF}}\leq \phi$. Here, $\phi$ is a target ratio that the miner can choose. E.g., if $\phi=2$, the miner weighs the probability of accepting the optimal transactions twice more than accepting an $\alpha$ fraction of zero-fee transactions.

\section{\rftm: Fairness in Transaction Fees Mechanism using Randomization}\label{sec::rtfm}
We now propose \rftm: a TFM that uses trusted on-chain randomness to guarantee both our fairness constraints, namely (i) \emph{ZTi} (Zero-Fee Transaction inclusion) and (ii) \emph{Monotonicity}. 
In addition to this, the proposed \rftm\ is both Miner Incentive Compatible (MIC) and Dominant Strategy Incentive Compatible (UIC).

We next (i) introduce randomized Transaction Fees Inclusion \rftm, (ii) show that when paired with the payment rules of FPA and EIP-1559, preserves their incentive guarantees while simultaneously satisfying the fairness notions (i) Monotonicity and (ii) ZTi.

\subsection{\rftm: Randomized TFM}\label{ssec::rtfm}


We denote \rftm\ as the tuple $\mathcal{T}^{\rftm}_{\phi} = \left(\bf{x}^{\rftm}_{\phi},\bf{p},\bf{q},\tau_{R}\right)$. At its core, \rftm\ comprises a novel allocation rule, $\bf{x}^{\rftm}$, and can be paired with any payment and burning rule. The allocation rule uses two sub-procedures: (i) \emph{transaction sampling} and (ii) \emph{biased coin-toss}. We first introduce these procedures and subsequently use them to formally define $\bf{x}^{\rftm}_{\phi}$.


\paragraph{Transaction Sampling.} An honest miner of a block adds transactions from the mempool $M$ to its block using the following rules.
\begin{itemize}[leftmargin=*]
    \item \underline{\textsc{Rule 1}}: The miner uniformly adds transactions from the mempool $M$ to its block $B_k$. But, for each transaction $t \in B_k$, the miner receives \underline{zero fees}. That is, $\forall t \in B_k, p_t=0$. Denote the Merkle tree constructed using these transactions as $\textsf{MT}_{\texttt{rand}}$ with the Merkle root, $\textsf{root}_{\texttt{rand}}$. 
    \item \underline{\textsc{Rule 2}}: The miner selects the transactions optimally, i.e., using Eq.~\ref{eqn::miner_opt}. Denote the Merkle tree constructed using these transactions according to $\textsf{MT}_{\texttt{opt}}$ with the Merkle root, $\textsf{root}_{\texttt{opt}}$. 
\end{itemize}

While mining a block, the miner selects transactions and constructs Merkle trees according to Rule 1 and Rule 2. Denote the transaction selection rule, given $M$, 
be represented as $\textsc{sample}(M) := \left((\textsf{root}_{\texttt{rand}}, \textsf{MT}_{\texttt{rand}}),(\textsf{root}_{\texttt{opt}}, \textsf{MT}_{\texttt{opt}})\right)$.

\paragraph{Trusted Biased Coin Toss.} \rftm's allocation rule selects one out of the two sets of transactions created from Rules 1 and 2. We now introduce an \textit{on-chain-based} biased coin toss method to select between the two sets. Let $\phi \in [0,1]$ denote the probability of heads (or $0$) and $1-\phi$ denote the probability of tails (or $1$). 

From Section~\ref{sec::preli}, a miner mines its block $B_k$ at height $k$ using the hash of the parent block $\textsc{Hash}(B_{k-1})$, the random nonce $\texttt{rand}$, the block height $k$, the two Merkle roots $\textsf{root}_{\texttt{rand}}$ and $\textsf{root}_{\texttt{opt}}$. If the block is mined, i.e.,  $\textsc{Hash}(B_{k}) < TD$ for target difficulty $TD$, then the toss' outcome is considered as follows:
\begin{equation}\label{eqn::coin-toss}
    O\left(\textsc{Hash}(B_{k}),\phi\right) := \begin{cases}
        0 & \text{if } \textsc{Hash}(B_{k}) < \phi\cdot TD \\
        1 & \text{otherwise}
    \end{cases}
\end{equation}

Remark~\ref{rem:coin-flip-equiv} shows that Eq.~\ref{eqn::coin-toss} is equivalent to a biased coin toss; refer to Appendix~\ref{app::rem:coin-flip-equiv} for the formal proof.

\begin{remark}\label{rem:coin-flip-equiv}
    Invoking  $O(\textsc{Hash}(B_{k}),\phi)$ for a mined block $B_{k}$ is equivalent to a biased coin toss with $\phi$ as the probability of heads.
\end{remark}

Given this, Algorithm~\ref{algo::rtfm} provides the procedural outline of $\bf{x}^{\rftm}_{\phi}$. The procedure is summarized as follows: 
\begin{itemize}[leftmargin=*]
    \item \underline{\textsc{Step 1}.} Miner samples two Merkle trees $\textsf{MT}_{\texttt{rand}}$ and $\textsf{MT}_{\texttt{opt}}$ by invoking $\textsc{sample}(M)$ and includes both Merkle roots $\textsf{root}_{\texttt{rand}}$ and $\textsf{root}_{\texttt{opt}}$ in block header $B_{k}$.
    \item \underline{\textsc{Step 2}.} Miner selects a random nonce for the block header $B_{k}$ until the block is mined; i.e. $\textsc{Hash}(B_{k}) < TD$. 
    \item \underline{\textsc{Step 3}.} Miner invokes biased coin toss $O(\textsc{Hash}(B_{k}),\phi)$ (Equation~\ref{eqn::coin-toss}). If the outcome is $1$, then $ \textsf{MT}_{\texttt{opt}}$ (optimally selected transactions) is considered part of the blockchain. If the outcome is $0$, then $\textsf{MT}_{\texttt{rand}}$  (Merkle tree with uniformly sampled transactions) is considered part of the blockchain.
\end{itemize}

To summarize, Definition~\ref{def::rtfm-alloc} formally defines $\bf{x}^{\rftm}$. 

\begin{definition}[\rftm\ Allocation Rule]\label{def::rtfm-alloc}
    Given $\mathcal{H}, M$ and $B_{k}$, let $x^{\rftm}_{\phi}$ denote a feasible allocation rule generated using Algorithm~\ref{algo::rtfm}. Formally, the set of allocated transactions $x^{\rftm}(\mathcal{H},M,B_{k},C,\phi) = \mathcal{X}_{k}$ for block $B_{k}$ is obtained from $(\mathcal{X}_{k},B_{k}) \gets \textsc{MineBlock}(C,M,p,\mathcal{H})$.
\end{definition}

\paragraph{rTFM Payment and Burning Rule.} The allocation rule $\bf{x}^{\rftm}_{\phi}$ can be coupled with any payment $(\bf{p})$ and burning $(\bf{q})$ rules to define $\mathcal{T}^{\rftm}_{\phi}$. E.g., similar to FPA, we can create $\mathcal{T}^{\rftm}_{\phi}$ such that each bidding user $i$ whose $t_{i} \in \mathcal{X}_{k}$ for $(\mathcal{X}_{k},B_{k}) \gets \textsc{MineBlock}(C,M,p,\mathcal{H})$ has $p_{i}^{FPA} = b_{i}$ else $p_{i}^{FPA} = 0$. In both cases, $q_{i}^{\rftm} = 0$. 

\begin{algorithm}[!t] 
\caption{\label{algo::rtfm} Randomized TFM (\rftm) Allocation Rule}
\begin{algorithmic}[1]
\Statex\textbf{Input:} Block Size $C$, Mempool $M$, Zero-Fees probability $\phi$, parent Block $B_{k-1}$, Target difficulty $TD$
\Statex\textbf{Output:} $(\textsf{MT}_{k},B_{k})$ Merkle Tree $\textsf{MT}_{k}$ of selected transactions and Mined block $B_k$
\Procedure{mineBlock}{$C,M,\phi,B_{k-1}$}
\State $\left((\textsf{root}_{\texttt{rand}}, \textsf{MT}_{\texttt{rand}}),(\textsf{root}_{\texttt{opt}}, \textsf{MT}_{\texttt{opt}})\right) \gets \textsc{sample}(M)$
\State $r \gets \textsc{random}()$ 
\Comment{\textcolor{blue}{Select a random nonce}}
\State $B_{k} \gets (B_{k-1},\textsf{root}_{\texttt{rand}},\textsf{root}_{\texttt{opt}},r)$ 
\Comment{\textcolor{blue}{Construct block $B_{k}$}}
\While{$\textsc{Hash}(B_{k}) \geq TD$}
    \State $r \gets \textsc{random}(\cdot)$
    \State $B_{k} \gets (B_{k-1},\textsf{root}_{\texttt{rand}},\textsf{root}_{\texttt{opt}},r)$ 
\EndWhile
\If{$\textsc{Hash}(B_{k}) \leq \phi\cdot TD$}\Comment{\textcolor{blue}{Biased coin-toss}}
\State \Return $(\textsf{MT}_{\texttt{rand}},B_{k})$
\Else 
\State \Return $(\textsf{MT}_{\texttt{opt}},B_{k})$
\EndIf
\EndProcedure

\end{algorithmic}
\end{algorithm}


\subsection{\rftm: Fairness Properties\label{subsec::rtfm_faitnot}} TFM with \rftm's allocation rule and EIP-1559's payment rule, i.e., $\mathcal{T}^{\rftm}_{p} = \left(\bf{x}^{\rftm},\bf{p}^{EIP-1559},\bf{q}^{EIP-1559},\tau_{R}\right)$ satisfies both Monotonicity and Zero-Fee Transaction Inclusion.  

\begin{theorem}\label{thm::rtfm::fairness}
  \rftm\ with EIP-1559 satisfies (i) Monotonicity and (ii) Zero-Fee Transaction Inclusion for any $\phi \in (0,1)$. 
\end{theorem}

Theorem~\ref{thm::imp} does not apply to \rftm\ as the miner does not have control over which set of transactions are selected with $\bf{x}^{\rftm}$. We can trivially extend Theorem~\ref{thm::rtfm::fairness} to show that \rftm\ with FPA also satisfies both fairness notions.

\subsection{\rftm\ : Incentive Properties}
We now discuss the incentive properties of $\mathcal{T}^{\rftm}_{\phi}$ with payment rules of (i) First Price Auction (FPA) and (ii) EIP-1559. First, we show that \rftm\  satisfies MIC for both FPA and EIP-1559 payment rules. Moreover, \rftm\ is UIC when the payment rule is EIP-1559. Following this, we also show that \rftm\  satisfies fairness properties, namely (1) \emph{ZTi} and (2) \emph{Monotonicity}.

\paragraph{\rftm\ with FPA.} The payment rule for FPA for any selected transaction $t_{i}$ with bid ${b}_{i}$ is ${p}_{i}^{FPA} = {b}_{i}$ if $t_{i} \in B_{k}$ and ${p}_{i}^{FPA} = 0$ otherwise. In both cases, the burning rule is ${q}_{i}^{FPA} = 0$. Trivially, \rftm\ with FPA is not UIC, while Theorem~\ref{thm::rtfm-mic} proves that it satisfies MIC. 

\paragraph{\rftm\ with EIP-1559.}  The EIP-1559 payment rule implies that for each bidding user $i$ whose $t_{i} \in B_{k}$ and $b_{i} \neq 0$ has ${p}_{i}^{EIP-1559} = b_{i} - \lambda$ and ${q}_{i}^{EIP-1559} = \lambda$. Here, $\lambda$ is the posted price determined by the network (refer to Footnote 6). With this, Theorem~\ref{thm::rtfm-dsic} shows that \rftm\ with EIP-1559 is UIC. 

\begin{theorem}\label{thm::rtfm-dsic}
     \rftm\ with EIP-1559's payment rule satisfies \emph{Dominant Strategy Incentive Compatibility} (UIC), if $\lambda$ is excessively low. 
\end{theorem}

Theorem~\ref{thm::rtfm-dsic} follows by observing that a user's strategy does not depend on \rftm's allocation but only on the payment and the burning rule. Thus, the UIC guarantee of EIP-1559 carries over for \rftm\ with EIP-1559. Furthermore, unlike \stfm\ (Section~\ref{subsec::soft_stfm}), \rftm\ is also MIC, as shown in Theorem~\ref{thm::rtfm-mic}.

\begin{theorem}\label{thm::rtfm-mic}
    \rftm\  satisfies Miner Incentive Compatibility (MIC) when transaction allocation rule is $\bf{x}^{\rftm}_{\phi}$ and payment scheme $\bf{p}^{\rftm}$ and burning rule $\bf{q}^{\rftm}$ are either (1) First Price Auction or, (2) EIP-1559.
\end{theorem}




\subsection{\rftm: Choosing $\phi$}
\rftm's allocation rule is parameterized by the probability $\phi$ of mining a block where each transaction $t_{i}$ has bid $b_{i} = 0$. We now discuss the impact of $\phi$ on \textsf{CoF} and the variation in the miner's revenue.

\paragraph{Cost of Fairness (CoF).} From Definition~\ref{def::cof}, \textsf{CoF} is the ratio of the utilities $u_{\texttt{opt}}$ (refer to Eq.~\ref{eqn::miner_opt}) and $u_{\rftm}$ (i.e., miner's utility when the transactions are selected according to $\bf{x}_{\phi}^{\rftm}$).

The miner's utility in \rftm\ is dependent on the output of random variable $O(\textsc{Hash}(B_{k}),\phi)$. If $O(\textsf{Hash}(B_{k}),\phi) = 0$ (occurs with probability $\phi$), then each selected transaction $t_{i}$ has $b_{i} = 0$ resulting in zero revenue for the miner. In contrast, with probability $1 - \phi$, we have $O(\textsc{Hash}(B_{k}),\phi) = 1$, such that the optimal transactions are selected. Here, the miner's revenue is equal to $u_{\texttt{opt}}$. That is, 
$$\mathbb{E}_{\phi}[u_{\rftm}] = \phi\cdot 0 + (1 - \phi)\cdot u_{\texttt{opt}}.$$

This implies that, $\textsf{CoF}_\rftm = \frac{u_{\texttt{opt}}}{\mathbb{E}_{\phi}[u_{\rftm}] } = \frac{1}{1 - \phi}$.

\smallskip\noindent\emph{\underline{Impact of $\phi$ on \textsf{CoF}}.} Trivially, an increase in $\phi$ increases ZTi. On the other hand, this also increases \textsf{CoF}, reducing the miner's revenue. However, since \rftm\ (with an appropriate payment rule) is MIC, we believe that the system designers must choose an appropriate $\phi$ which (i) incentivizes the miner to not abstain from the system and (ii) allows for a desirable percentage of zero-fee transactions that may lead to greater adoption. 

\paragraph{Coefficient of Variation (CoV).} An increase in $\phi$ not only decreases the miner's expected revenue but will also \textit{increase} its variance. More concretely, denote $\sigma_{\texttt{opt}}$ as the standard deviation and $\pi_{\texttt{opt}}$ as the miner's expected utility when it optimally selects the transactions. Likewise, $\sigma_{\rftm} \mbox{~and~} \pi_{\rftm}$ are the standard deviation and expectation in the miner's utility from \rftm.  We know that the Coefficient of Variation (CoV) is given by $\frac{\sigma}{\mu}$. By trivial arguments, we know the following: 
$$
CoV_{\texttt{opt}} = \frac{\sigma_{\texttt{opt}}}{u_{\texttt{opt}}} = 1 \mbox{~\textbf{\&}~} CoV_{\rftm} = \frac{\sigma_{\rftm}}{\mathbb{E}_\phi[u_{\rftm}]} = \left(\frac{1 - \phi}{\phi}\right)^{1/2}
$$
Ideally, we want to choose $\phi$ such that ${CoV^{2}_{\texttt{opt}}}/{CoV^{2}_{\rftm}}$ is maximized. Towards this, we observe that as $\phi \to 0$, the CoV ratio increases monotonically.

\begin{figure}
    \centering
    \includegraphics[width=\columnwidth,trim={150pt 100pt 260pt 30pt},clip]{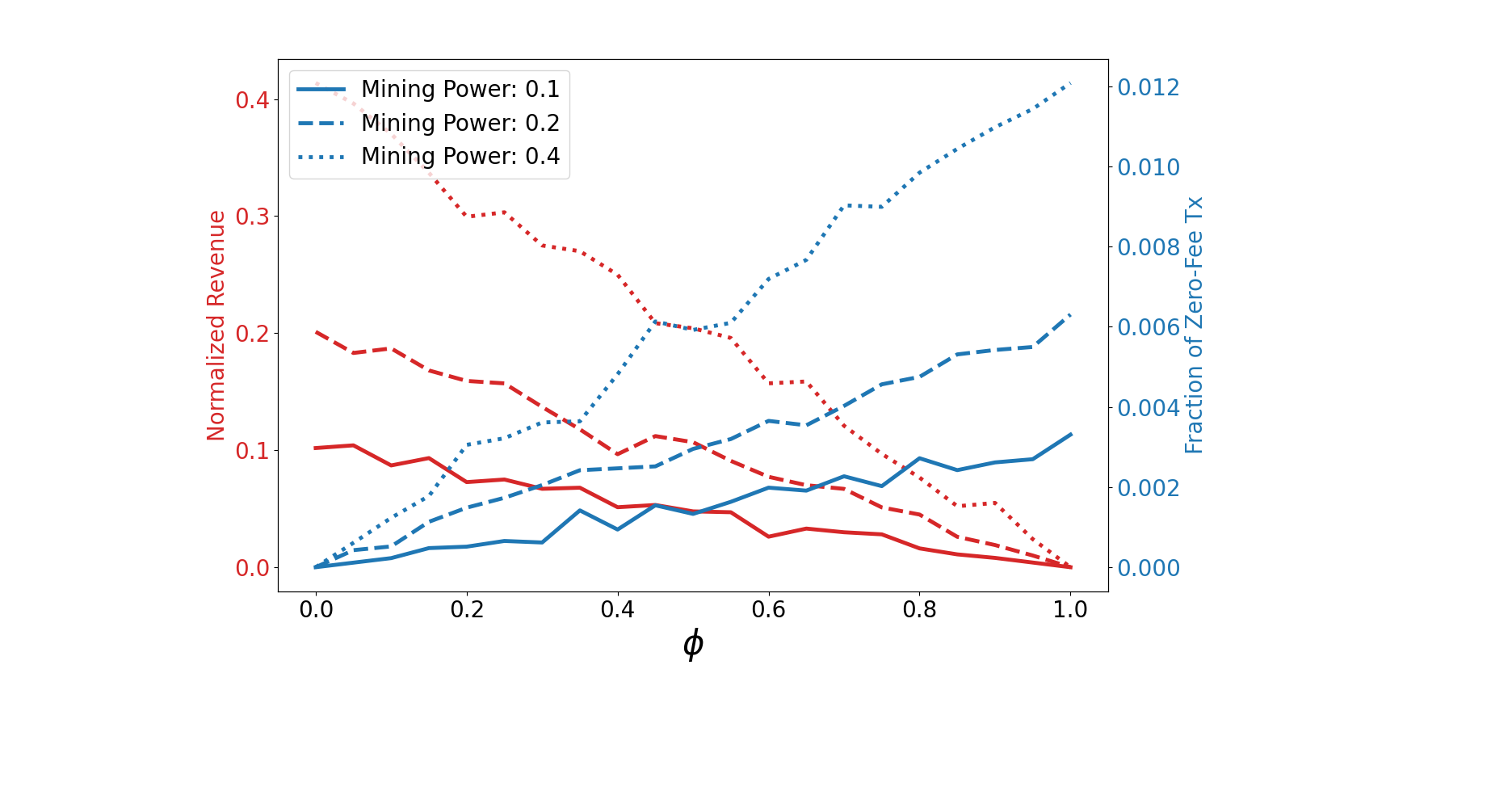}
    \caption{\rftm: Effect of $\phi$}
    \label{fig:my_phi}
\end{figure}

\subsubsection{\rftm: Empirical Analysis}

\paragraph{Setup.} To simulate \rftm, we need to configure the size of the mempool $M$, block size $C$, $\phi$, sample each user's bid $b$, and their sizes. Our experiments consider transactions of the same size $(s_i=s_j=1)$. We set the mempool size as $n = 1000$, block size $C = 100$, and each user's bid is sampled from \emph{Truncated Gaussian} distribution, $b \sim \mathcal{N}(4,3)$.

\paragraph{Measures.} We vary $\phi\in [0,1]$ and observe the (i) \textit{Normalized Miner Revenue}, ratio of miner's revenue from \rftm\ with \texttt{OPT} and (ii) \textit{Fraction of Zero-fee Txs}, ratio of zero-fee transactions accepted in \rftm\ with the mempool size. We report the average results across $1000$ runs. 

\paragraph{Results.} Figure~\ref{fig:my_phi} depicts our results. As expected, an increase in $\phi$ increases the zero-fee transactions included and decreases the miner's revenue. In the Appendix, we also show that the trends depicted in Figure~\ref{fig:my_phi} remain the same when $b\sim \mbox{U}[0,1]$ and $b\sim \mbox{Exp}(\lambda = 1.5)$.

\paragraph{Choosing $\phi$.} In summary, the trade-off between (1) \textsf{CoF}, (2) CoV, and (3) Fraction of Zero-fee Txs is such that as $\phi$ increases, \textsf{CoF} increases, CoV-ratio decreases, and ZTi increases. If we wish to increase the number of accepted zero-fee transactions, we must compromise with utility and suffer higher variance. Figure~\ref{fig:my_phi} depicts the said trade-off empirically.

\paragraph{Discussion.}  In summary, both \stfm\ and \rftm\ satisfy ZTi and Monotonicity, with \rftm\ also being MIC (for an appropriate payment rule). However, \stfm\ is relatively simpler to implement since miners must only adapt to the new allocation rule. In contrast, \rftm\ will require a fork of the blockchain. The coin-toss mechanism introduced for \rftm\ also requires a PoW blockchain, which is often resource-intensive. Future work can extend \rftm's allocation rule for other blockchains (e.g., Proof-of-Stake blockchains).

\section{Conclusion}
In this paper, we focused on the need for fairness in TFMs regarding the transaction fees for the transaction creators. We argued that including zero-fee transactions is necessary for the widespread adoption of TFMs. We introduced two novel fairness notions:  Zero-fee Transaction Inclusion (ZTi) and Monotonicity. We showed that existing TFMs do not satisfy at least one of these notions or do so for smaller transaction sizes and at a high cost to the miner's utility. To resolve these limitations, we first introduced \stfm\, which samples transactions through the distribution generated from the softmax with temperature $(\gamma)$ function. We showed that while \stfm\ is a fair TFM, it is not MIC. To this end, we introduced \rftm\, which simultaneously satisfies MIC and our fairness notions. 

\paragraph{Future Work.} We believe that these fair TFMs may further democratize TFMs by contributing to their broader accessibility and enhancing their adoption in the market. Future work can further study the role of $\phi$ in \rftm\ towards striking a desirable balance between a miner's revenue and the fraction of zero-fee transactions included. Last, as aforementioned, one can also explore extending \rftm\ for Proof-of-Stake blockchains.

\bibliographystyle{ACM-Reference-Format} 
\bibliography{ref}


\appendix

\newpage
\section{Proofs}

\subsection{Proof of Remark 1\label{remark1}}

\begin{proof}
Consider the following example, where each transaction is of the same size. Let $n=5$ such that the current block $B_k$ can hold up to $8$ transaction. Further, we have $\alpha=3/4$. The miner must add (any) $2$ transactions to the $1-\alpha$ section first before greedily adding transactions to the $\alpha$ section. Whichever transactions from $M$ the miner chooses to add to the $1-\alpha$ section, it can strictly increase its utility by adding $2$ fake transactions instead. That is, by adding these fake transactions, the miner can add the real transactions of $M$ to the $\alpha$ section. Thus, BitcoinF's allocation rule does not satisfy MIC.
\end{proof}

\subsection{Proof of Claim~\ref{claim::BF_bound}}\label{app:claim-BF-bound}

\begin{proof}
W.l.o.g., let the optimal set of bids (sorted in non-decreasing order) which maximizes the miner's utility in Eq.~\ref{eqn::miner_opt} with $p_t=b_t$ and $q_t=0,~\forall t$ be $\{b_1,\ldots,b_c\}$. Then with $\alpha=\frac{k}{c}$ s.t. $k\leq c$, we can write BitcoinZF's bid set as $\{b_1,\ldots,b_k\}$ (since the miner will maximize utility in the "$\alpha$" section of the block). Observe that,
\begin{equation*}
    \begin{aligned}
    \frac{OPT}{u_m^{BZ}} & = \frac{b_1+\ldots+b_c}{b_1+\ldots+b_k} = 1 + \frac{b_{c-k+1}+\ldots+b_c}{b_1+\ldots+b_k} \\
    & \leq 1 + \frac{(c-k)b_k}{k\cdot b_k} \leq 1 + \frac{c}{k} -1 \leq \frac{c}{k} = 1/\alpha.
    \end{aligned}
\end{equation*}
This completes the claim.
\end{proof}

\subsection{Proof of Theorem~\ref{thm::imp}}\label{app:thm-imp}

\begin{proof}
Consider the following example. Let the transaction bid and size pair in the mempool be denoted by $\mathcal{P}=[(b_i,s_i)]=\{(10,10),(10,10),(5,10),(0,10),(0,10)\}$. If the block $B_k$ can admit a total transaction size of $30$, then the miner can maximize its utility from~\ref{eqn::miner_opt} by selecting the first three transactions in $\mathcal{P}$. That is, $\mathbf{x}^{TFM}=\{1,1,1,0,0\}$ with $u_m^{TFM}=25$. This implies that $\Pr(t_4\in B_k)=\Pr(t_5\in B_k)=0$, thus, ZTi is not satisfied. 
\end{proof}

\subsection{Proof of Theorem~\ref{thm::BF-Fairness}}\label{app:thm-BF-Fairness}

Without considering the inclusion of fake bids from the miner, i.e., $F=\emptyset$,  we first write the optimization of BitcoinZF as follows:

\begin{align}\label{eqn::new_BF_opt_1}\tag{B1}
\begin{rcases}
    \max_{\mathbf{x}^{BZ}} & \sum_{i\in M} x_i^{BZ}\cdot p_i^{BZ}(\mathcal{H},B_k)\cdot s_i\\
    \mbox{s.t.~} & \sum_{t\in M, b_t\not=0}s_t\cdot x_t^{BZ}(\mathcal{H},M) \leq C_{\alpha} \\
    & \sum_{t\in M, b_t=0}s_t\cdot x_t^{BZ}(\mathcal{H},M) = C_{1-\alpha} \mbox{~and~}  \\
   & x_t^{BZ}(\mathcal{H},M)\in\{0,1\}, \forall t\in M.
\end{rcases}
    \end{align}

To show that BitcoinZF satisfies Monotonicity, we have to show that by increasing its bid $b_i$, agent $i$'s transaction $t_i$ has a higher probability of getting accepted in $B_k$. Indeed, this is the case in BitcoinZF, since increasing $b_i$ to $b_i+\epsilon$ s.t. $\epsilon>0$, can only increase the probability of $t_i$'s inclusion in $B_k$. This is because of the KNAPSACK definition from Eq.~\ref{eqn::new_BF_opt_1}.

Furthermore, since the miner receives no utility from any transaction in the ``$1-\alpha$" section, it can uniformly sample zero-fee transactions in this section. There is a subtle point here: the miner does not get any utility by adding these transactions to the $1-\alpha$ section. It can, in effect, leave the section empty or add its own transactions. However, since such deviations will not yield the miner any increase in utility, we can state that BitcoinZF satisfies ZTi.

\subsection{Proof of Theorem~\ref{thm::sftm_fair}}\label{app:thm-sftm-fair}

\begin{proof}
We first prove that STFM satisfies Monotonicity irrespective of the payment and burning rules.

For this, we must show that $\forall t_i,t_j\in M$ s.t. $t_i\not=t_j$ if $b_i>b_j$, $\Pr(t_i\in B_k)>\Pr(t_j\in B_k)$. We remark that $\mathbf{x}^{STFM}$ admits transactions with a distribution generated by applying the softmax function on the transactions in $M$ (refer Algorithm~\ref{algo::sftm}). 

For sampling the first transaction, the probability distribution is $\Pr(t_i\in B_k)=\frac{\exp(b_i/\gamma)}{\sum_{i^\prime\in M}\exp(b_{i^\prime}/\gamma)},~\forall i \in M$. Trivially, we have $\frac{\exp(b_i/\gamma)}{\sum_{i^\prime}\exp(b_{i^\prime}/\gamma)}>\frac{\exp(b_j/\gamma)}{\sum_{i^\prime}\exp(b_{i^\prime}/\gamma)}$ if $b_i>b_j$ and $\gamma> 0$, implying STFM satisfies Monotonicity in this case. Next, w.l.o.g., we assume a transaction $t_l$ was sampled. The re-generated probability distribution becomes, $\Pr(t_i\in B_k)=\frac{\exp(b_i/\gamma)}{\sum_{i^\prime\in M\setminus \{l\}}\exp(b_{i^\prime}/\gamma)},~\forall i \in M\setminus \{l\}$. Still, we have $\frac{\exp(b_i/\gamma)}{\sum_{i^\prime}\exp(b_{i^\prime\in M\setminus \{l\}}/\gamma)}>\frac{\exp(b_j/\gamma)}{\sum_{i^\prime}\exp(b_{i^\prime\in M\setminus \{l\}}/\gamma)}$ if $b_i>b_j$ and $\gamma> 0$. That is, Monotonicity still holds. Along similar lines, we can show that Monotonicity holds for each sampling stage. 

Trivially, we can also show that STFM satisfies Zero-fee Transaction Inclusion (ZTi). For each $t_i\in M$ with $b_i=0$, we have $\Pr(t_i\in B_k) = \frac{\exp(b_i/\gamma)}{\sum_{i^\prime\in M}\exp(b_{i^\prime}/\gamma)} = \frac{1}{\sum_{i^\prime\in M}\exp(b_{i^\prime}/\gamma)}>0$, irrespective of the size of $M$. 
 
\end{proof}

\subsection{Proof of Theorem~\ref{thm::sftm-mic-general}}\label{app:thm-sftm-mic-general}

\begin{proof}
For the proof, we have to show that for any non-trivial payment rule, the intended allocation $\mathbf{x}^{STFM}$ in $\mathcal{T}^{STFM}$ is such that the miner has an incentive to deviate.

Given the mempool $M$, denote $Z\subset M$ as the set of all zero-fee transactions, i.e., $Z=\{t_i~|~t_i\in M~\mbox{~and~}b_i=0\}$. For all game instances of $\mathcal{T}^{STFM}$  where the block $B_k$'s size $C$ is less than the size of the transactions in $M-Z$, we have $\Pr(t_i\in B_k)=0,\forall t_i \in Z$. That is, the miner has no incentive to add transactions in $Z$ to $B_k$. This is because as the payment rule is increasing with the transaction fees, the miner's utility from greedily adding transactions from $M-Z$ will be strictly greater than including even a single transaction from $Z$. 
\end{proof}

\subsection{Proof of Theorem 5\label{thm4}}

\begin{proof}

Denote $C$ and $N$ as the block size and mempool size, respectively. Let OPT denote miner's utility from Eq.~\ref{eqn::miner_opt} with $p_t=b_t$ and $q_t=0,~\forall t$  and $u^{STFM}_m$ denote miner's utility for $\mathcal{T}^{STFM}$. Let $M$ comprise $n$ transactions with fees $\{b_{i}\}_{i=1}^{n}$. W.l.o.g, we consider $b_{1} \geq b_{2} \geq \ldots \geq b_{n}$. Let $c$ denote the maximum number transactions in a block. The block-size is $C$ and for simplification we assume that transactions are of the same size.\footnote{if $N$ and $C$ are large enough, then with very high probability, number of transactions $n$ (or $c$) in a pool (or block) of size $N$ (or $C$) will deviate from $n$ (or $c$) negligibly. This observation follows from Chernoff bound.}

 Miner's optimal utility from Eq.~\ref{eqn::miner_opt}  is:  $OPT = \sum_{i=1}^{c} b_{i}$.
Let $X$ denote the utility from sampling one transaction from $M$ using $\mathcal{T}^{STFM}$. Then, $\mathbb{E}[X] = \sum_{i=1}^{n} \Pr(t_i\in B_k)\cdot b_{i}$.  Further, if $X_{i}$ is the utility from $i^{th}$ sampled transaction, (out of total $c$ transactions present in a block), then the expected utility is given by
$$\mathbb{E}[u^{STFM}_m] = \mathbb{E}[\sum_{x=1}^{c}X_{x}] = |c|\sum_{i=1}^{n} b_{i}\Pr(t_i\in B_k).$$
We get the last equation using linearity of expectations. Therefore, the ratio of utilities is,
$$ 
\frac{OPT}{\mathbb{E}[u_m^{STFM}]}  = \frac{\sum_{i=1}^{c} b_{i}}{|c|\sum_{i=1}^{n} b_{i}\Pr(t_i\in B_k)}
$$
For maximizing $ \frac{OPT}{\mathbb{E}[u_m^{STFM}]}  $ we need to maximize numerator and minimize denominator. This is achieved by taking $b_{1} = b_{2} = \ldots = b_{c} = b$ and $b_{c+1} = b_{c+2} = \ldots = b_{n} = 0$. That is,
\begin{equation*}
\begin{aligned}
\frac{OPT}{\mathbb{E}[u_m^{STFM}]}   &= \frac{c\cdot b}{c(\sum_{i=1}^{c}b\cdot\;\frac{e^{\frac{b}{\gamma}}}{n - c + c\cdot\;e^{\frac{b}{\gamma}}} + 0)}\\
\frac{OPT}{\mathbb{E}[u_m^{STFM}]} &= \frac{n - c + c\cdot\;e^{\frac{b}{\gamma}}}{c\cdot\;e^{\frac{b}{\gamma}}} = \frac{n}{c} + 1 - e^{-\frac{b}{\gamma}}
\end{aligned}
\end{equation*} 
Upper bound on utility-loss $(\frac{OPT}{\mathbb{E}[u_m^{STFM}]})$ is found when $b \to \infty$ and is equal to $\frac{n}{C} + 1$.
\end{proof}

\subsection{Proof of Remark~\ref{rem:coin-flip-equiv}\label{app::rem:coin-flip-equiv}}
\begin{proof}
    From Eq.~\ref{eqn::coin-toss} we get $O(\textsc{Hash}(B_{k}),\phi) = 1$ if $\textsc{Hash}(B_{k}) < \phi\cdot TD$. For any hash function $\textsc{Hash}:\{0,1\}^{*}\rightarrow\{0,1\}^{\lambda}$, the pre-image guarantee implies $\textsc{Hash}(B_{k}) \in_{R} \{0,1\}^{\lambda}$~\cite{Robshaw2011HashOWF}
. However, since we are considering invocation for a mined block, we have $\textsc{Hash}(B_{k}) \in_{R} \{0,1,\ldots,TD-1\}$. As such, $\Pr\left(\textsc{Hash}(B_{k}) < \phi\cdot TD\right) = \phi$. The outcome $O(\textsc{Hash}(B_{k}),\phi) = 1$ is with probability $\phi$. We, therefore, get the equivalence by mapping this outcome to ``$Heads$" in a biased coin toss.
\end{proof}

\subsection{Proof for Theorem~\ref{thm::rtfm::fairness}}\label{app:thm-rftm-fairness}
\begin{proof}
     We show the proof in two steps.
    \begin{itemize}[leftmargin=*]
        \item \textbf{Monotonicity.} To show that \rftm\ satisfies Monotonicity, we have to show that increasing the bid $b_t$ of an arbitrary transaction $t$ increases its probability of acceptance in \rftm\ (given the remaining bids are fixed). For this, first, let us write down the probability of any transaction $t\in M$ with bid $b_t$ getting added to the block $B_k$. We have,
        \begin{equation}\label{eqn::t_added_to_Bk}
            \Pr(t\in B_k) = \phi \cdot \Pr(t\in MT_{opt}) + (1-\phi) \cdot \Pr(t\in MT_{\textsf{rand}})
        \end{equation}
        Now, assume that the new bid is $b_t + \epsilon$ for any $\epsilon > 0$. If $\Pr(t\in B_k)$ increases in Eq.~\ref{eqn::t_added_to_Bk} for $b_t + \epsilon$ (compared to the bid $b_t$), \rftm\ satisfies Monotonicity. Note that the term $\Pr(t\in MT_{\textsf{rand}}$ remains the same for both bids $b_t$ and $b_t + \epsilon$ as the miner does the allocation uniformly. The term $\Pr(t\in MT_{opt})$ can only increase for $b_t + \epsilon$ compared to $b_t$, for some $\epsilon$. This is because, in $MT_{opt}$, the miner adds the transactions optimally, i.e., using Eq.~\ref{eqn::miner_opt}. That is, \rftm\ satisfies Monotonicity.

        \item \textbf{Zero-fee Transaction Inclusion.} For any $\phi \in (0,1)$ and Eq.~\ref{eqn::t_added_to_Bk}, the probability of any transaction $t$ with $b_t = 0$ being part of the block $B_k$ is trivially non-zero. This is because $\Pr(t\in MT_{\textsf{rand}}$ will be non-zero even if $b_t = 0$.
    \end{itemize}
    These two steps complete the proof of the theorem.
\end{proof}

\subsection{Proof of Theorem~\ref{thm::rtfm-dsic}}
\begin{proof}
    Theorem~\ref{thm::rtfm-dsic} follows trivially by observing that an agent's strategy does not depend on \rftm's allocation but only on the payment and the burning rule. The TFM also satisfies Monotonicity (Theorem~\ref{thm::rtfm::fairness}). Thus, the UIC guarantee of EIP-1559 carries over for \rftm\ with EIP-1559. 
\end{proof}

\subsection{Proof for Theorem~\ref{thm::rtfm-mic}}\label{app::thm-rftm-mic}
\begin{proof}
To show that the TFMs satisfy MIC, we remark that the selecting between the optimal and zero-fee transactions (refer Algorithm~\ref{algo::rtfm}) is carried out by the blockchain in a trusted manner~(Eq.~\ref{eqn::coin-toss}). As the miner has no control over the random outcome of $O(\textsc{Hash}(B_{k}),\phi)$ (Remark~\ref{rem:coin-flip-equiv}), its strategy involves (i) optimally selecting the transactions and (ii) either adding the zero-fee transactions or keeping them empty. For (i), we know that both EIP-1559 and FPA payment rules satisfy MIC. For (ii), both strategies result in zero utility for the miner; that is, \rftm\ is MIC for the miner.
\end{proof}

\begin{figure*}[t]
    \centering
    \includegraphics[width=\linewidth]{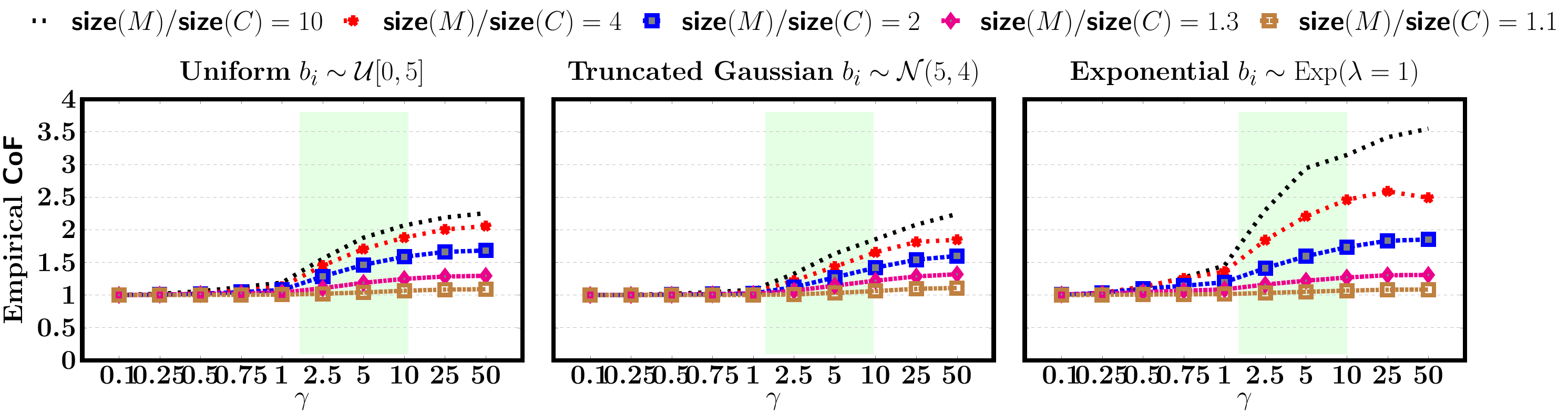}
     \renewcommand{\thefigure}{D1}
    \caption{Empirical \textsf{CoF} for the distributions: (\textsf{D1}) Uniform, (\textsf{D2}) Truncated Gaussian and (\textsf{D3}) Exponential.}
    \label{fig:RL}
\end{figure*}
\begin{figure*}[t]
    \centering    \includegraphics[width=\linewidth]{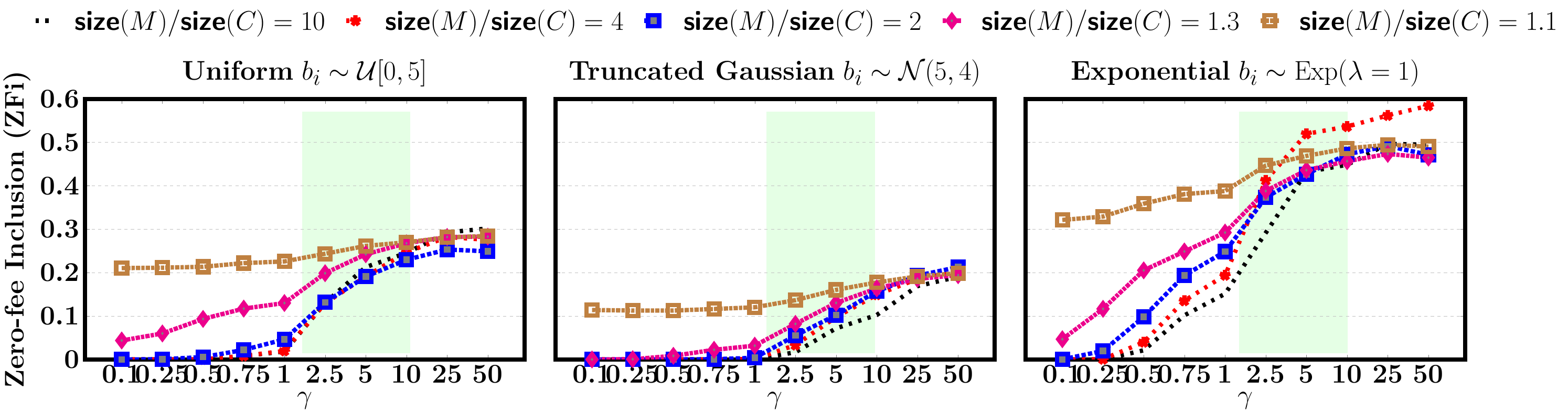}
     \renewcommand{\thefigure}{D2}
    \caption{Zero-fee Inclusion (ZFi) for the distributions: (\textsf{D1}) Uniform, (\textsf{D2}) Truncated Gaussian and (\textsf{D3}) Exponential.}
    \label{fig:zti}     
\end{figure*}

\begin{figure*}[th]
    \centering
    \includegraphics[width=\linewidth]{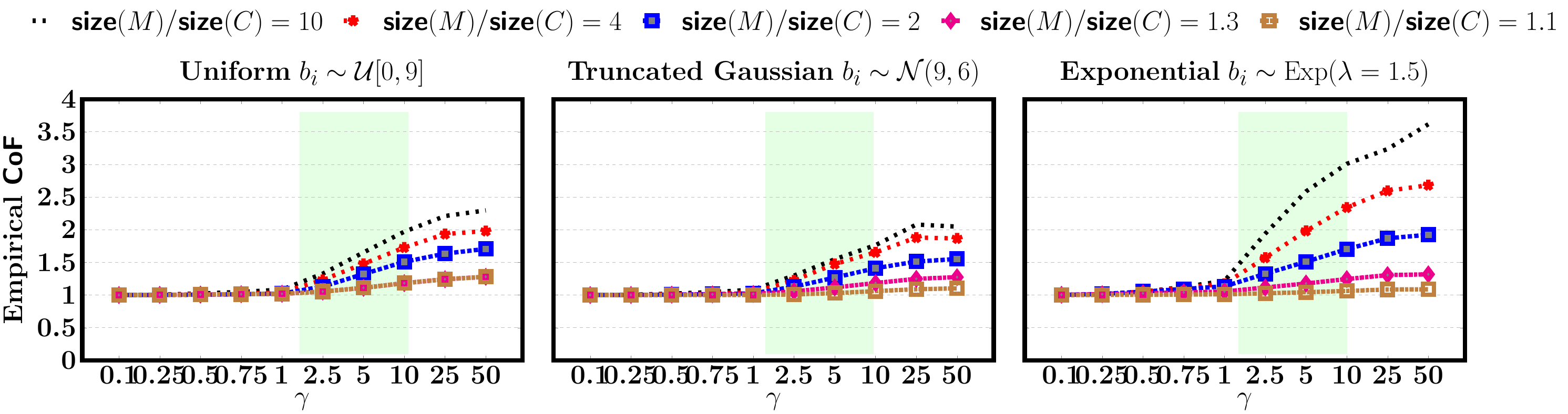}
 \renewcommand{\thefigure}{D3}
    \caption{Empricial \textsf{CoF}: Miner's Utility Ratio for the distributions: (\textsf{D1}) Uniform, (\textsf{D2}) Truncated Gaussian and (\textsf{D3}) Exponential}
    \label{fig:RL2}
\end{figure*}
\begin{figure*}[th]
    \centering
    \includegraphics[width=\linewidth]{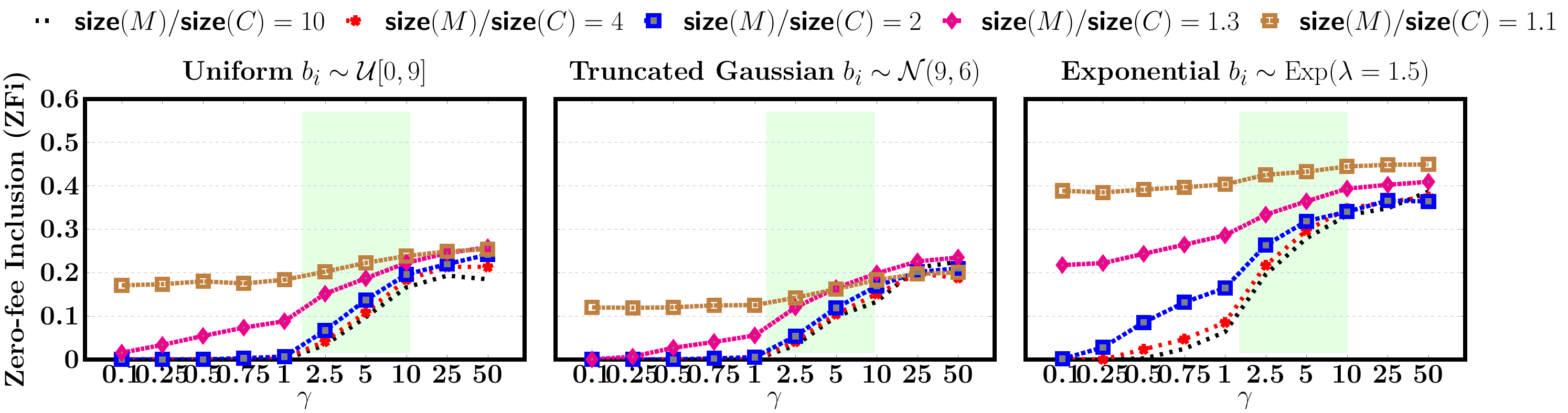}
    \renewcommand{\thefigure}{D4}
    \caption{Zero-fee Inclusion (ZFi) for the distributions: (\textsf{D1}) Uniform, (\textsf{D2}) Truncated Gaussian and (\textsf{D3}) Exponential}
    \label{fig:zti2}     
\end{figure*}

\section{Simulations}\label{sec::simulation}

We now empirically validate STFM's performance with regard to the loss in the miner's utility and the fraction of zero-fee transactions included in the block. 

\subsection{Experimental Setup \& Performance Measures}
\label{ssec::experimental-setup}
 To simulate STFM, we need to configure the size of the mempool $M$, block size $C$, temperature parameter $\gamma$, each agent's transaction fees, and their sizes. In our experiments, we vary the ratio of the sizes of the mempool and block size, say $\frac{\textsf{size}(M)}{\textsf{size}(C)}$, in the set $\{1.1,1.3,2,4,10\}$ and $\gamma\in[0.1,50]$. To concretely mimic all possible real-world scenarios, for each $t_i\in M$, the agent $i$ samples its bid $b_i$ from the following three distributions\footnote{We observe similar trends for other distribution parameters.}: (\textsf{D1}) Uniform, i.e., $b_i\sim \mathcal{U}[0,5]$, (\textsf{D2}) Truncated Gaussian, i.e., $b_i\sim \mathcal{N}(5,4)$, and (\textsf{D3}) Exponential, i.e., $b_i\sim \mbox{Exp}(\lambda=1)$.

Likewise, for each $t_i\in M$, the agent $i$ samples the transaction's size $s_i\sim \mbox{Exp}(\lambda=1)$. This choice is reasonable since smaller transactions (e.g., payer-payee token transfer) are more common than larger transactions (e.g., smart contract deployment). To measure STFM's performance, we also define the following measures. 
\begin{enumerate}[leftmargin=*,noitemsep]
    \item \underline{Empirical \textsf{CoF}}. This is the ratio of the miner's utility by greedily adding transactions to the block with the utility from STFM's allocation. The smaller the \textsf{CoF}, the better.
    \item \underline{Zero-fee Inclusion (ZFi)}. ZFi is the ratio of the size of zero-fee transactions in the block with the total size of all the transactions in the block.
\end{enumerate}

For each $\frac{\textsf{size}(M)}{\textsf{size}(C)}$ and $\gamma$, we sample the agent's bids based on \textsf{D1}, \textsf{D2} and \textsf{D3}. We simulate the resulting game instances 100 times and report the average \textsf{CoF} and ZFi values. The codebase is available with the accompanying supplement.

\subsection{Results \& Discussion}
\label{ssec::results}
Figure~\ref{fig:RL} and Figure~\ref{fig:zti} depict our results. Details follow.

\paragraph{Empirical \textsf{CoF}: Miner's Utility Ratio.} We first discuss the change in \textsf{CoF} with varying $\gamma$s and $\frac{\textsf{size}(M)}{\textsf{size}(C)}$ values for \textsf{D1}, \textsf{D2} and \textsf{D3}. For all three distributions, we observe a consistent increase in \textsf{CoF} as $\gamma$ increases, i.e., $\gamma\uparrow\implies u_m\downarrow$. For $\gamma\in(0,1)$, \textsf{CoF} is $<1.5$ implying that miner's utility drop is $>0.67$ times OPT. For $\gamma\geq 1$ \textsf{CoF} increases, but remains $<2.5$ for \textsf{D1}, \textsf{D2} and $<3.5$ for \textsf{D3}. E.g., for $\gamma=5$ and the worst-case value of $\frac{\textsf{size}(M)}{\textsf{size}(C)}=10$, \textsf{CoF} values are $1.88$ (\textsf{D1}), $1.63$ (\textsf{D2}) and $2.93$ (\textsf{D3}). 

Furthermore, one way to interpret decreasing $\frac{\textsf{size}(M)}{\textsf{size}(C)}$ is an increase in the block size, $\textsf{size}(C)$. As ${\textsf{size}(C)}\to{\textsf{size}(M)}$,  the randomized allocation adopted with STFM plays a lesser role as the block gets large enough to accommodate most transactions. From Figure~\ref{fig:RL}, we see that decreasing $\frac{\textsf{size}(M)}{\textsf{size}(C)}$ decreases \textsf{CoF}, i.e., an increase in the miner's utility. 

\paragraph{Zero-fee Inclusion (ZFi).} We empirically show that STFM admits zero-fee transactions with Figure~\ref{fig:zti}. For varying $\gamma$, we plot the ratio of the size of zero-fees transactions included in the block with the total block size (aka ZFi). We make five major observations. First, for $\gamma\in(0,1]$ and high $\frac{\textsf{size}(M)}{\textsf{size}(C)}$, ZFi values are $\approx 0$. Second, ZFi consistently increases as $\gamma$ decreases. Third, for $\gamma>5$, ZFi values almost saturates at $\approx 0.3$ (\textsf{D1}, \textsf{D2}) and $\approx 0.6$ (\textsf{D3}) for all values of $\frac{\textsf{size}(M)}{\textsf{size}(C)}$. Fourth, as $\frac{\textsf{size}(M)}{\textsf{size}(C)}$ decreases, we observe significant ZFi even for $\gamma\in (0,1)$. This is because smaller $\frac{\textsf{size}(M)}{\textsf{size}(C)}$ implies enough room for most of the available transactions. Lastly, since with \textsf{D3}, there is a greater chance of sampling lower $b_i$s, its ZFi values are greater than \textsf{D1} and \textsf{D2}.

The green shaded region depicts the range of $\gamma$ with a practical \textsf{CoF}-ZFi trade-off. Specifically, for $\gamma\in (2,10)$, we observe $\textsf{CoF}<2$ and ZFi $>0.1$, for all three distributions.

\begin{figure*}
\centering
\begin{minipage}{.5\textwidth}
  \centering
  \includegraphics[width=\linewidth]{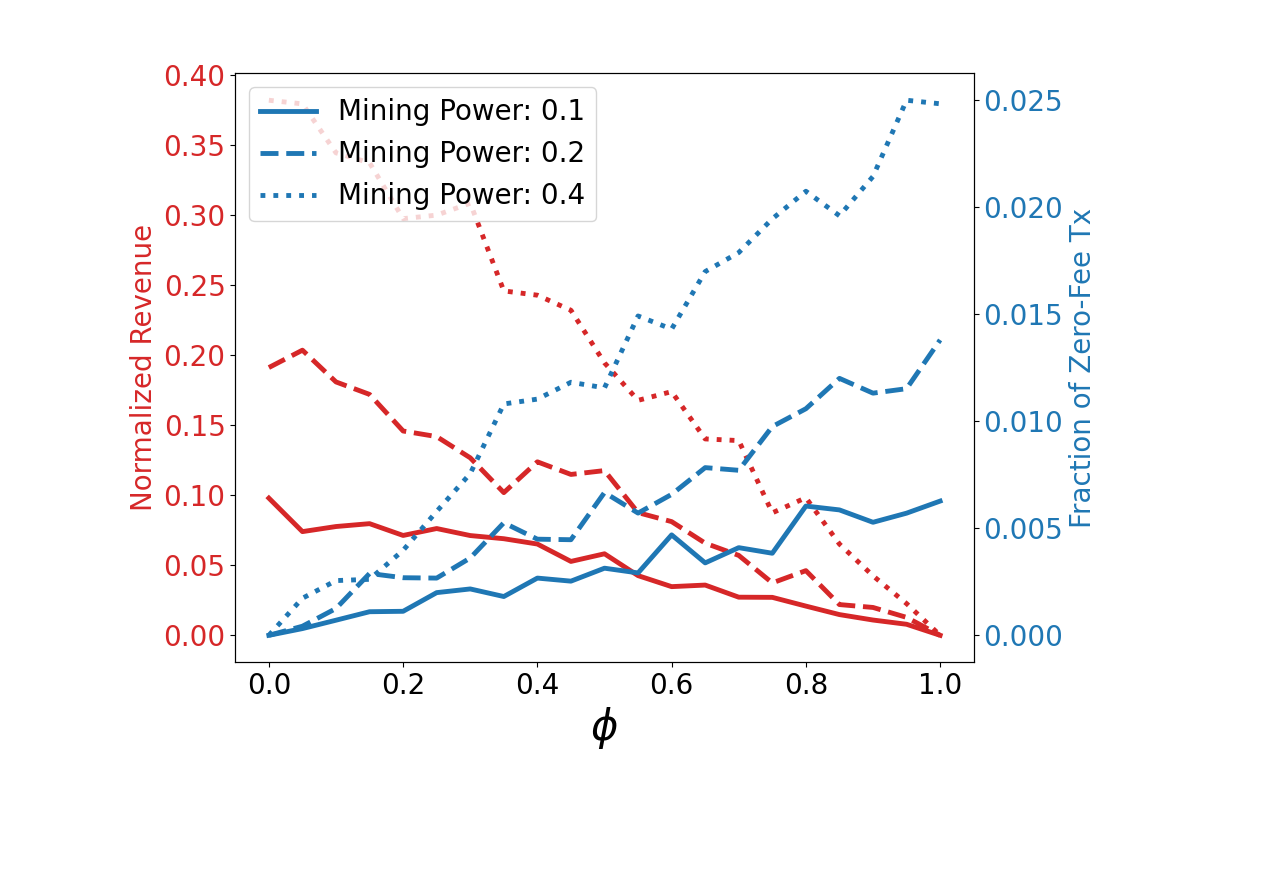}
       \renewcommand{\thefigure}{D5}
  \captionof{figure}{$b\sim\mbox{U}[0,1]$}
  \label{fig:test1}
\end{minipage}%
\begin{minipage}{.5\textwidth}
  \centering
  \includegraphics[width=\linewidth]{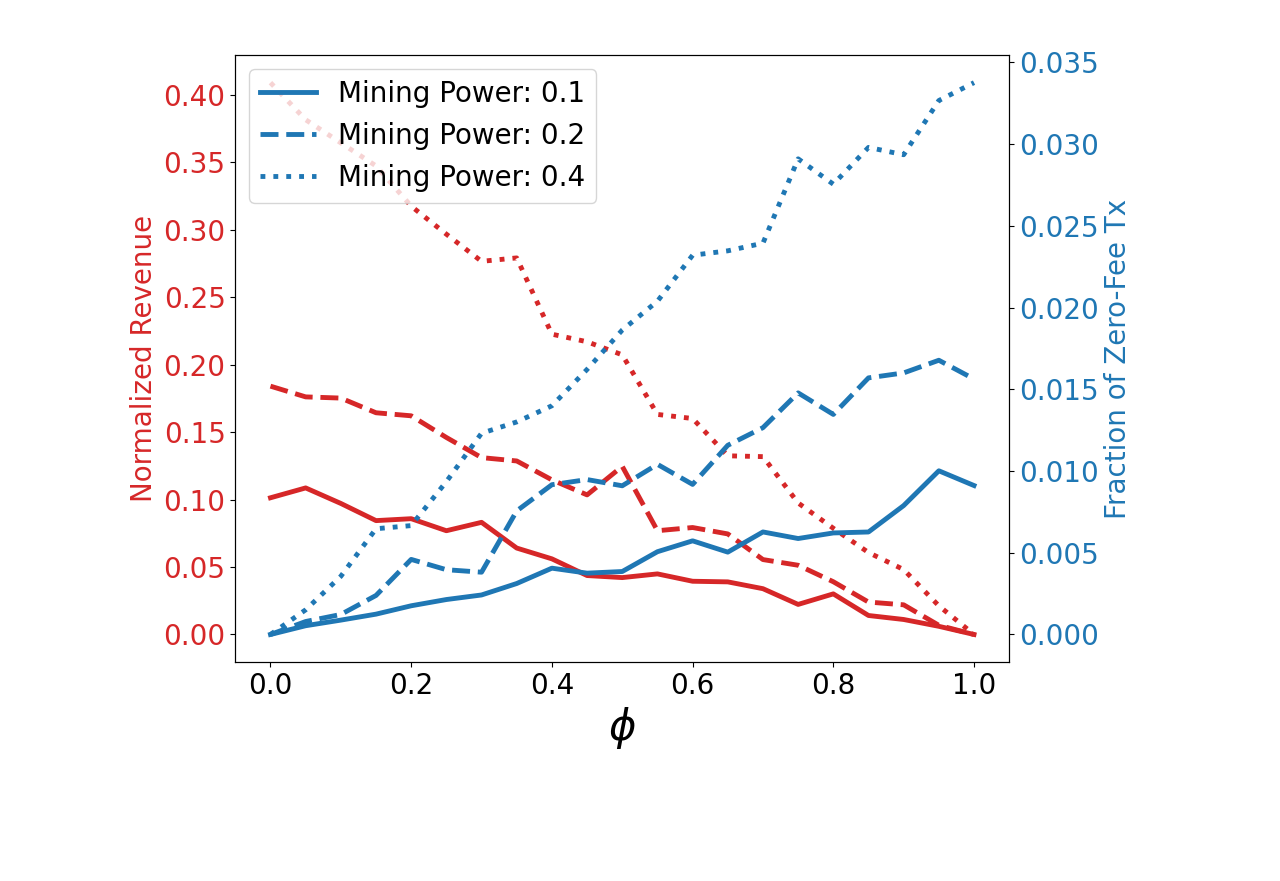}
       \renewcommand{\thefigure}{D6}
  \captionof{figure}{$b\sim\mbox{Exp}(\lambda=1.5)$}
  \label{fig:test2}
\end{minipage}
\end{figure*}

\subsection{\rftm: Additional Experiments}
In Section~\ref{sec::rtfm}, we presented experiential results when the bid distribution was from the Truncated Gaussian distribution. Here, we provide experiments for other bid distributions. Specifically, we provide results when the bid distributions are the Uniform and the Exponential distribution. That is, we have $b\sim\mbox{U}[0,1]$ and $b\sim\mbox{Exp}(\lambda=1.5)$. 

 Figure~\ref{fig:test1} and Figure~\ref{fig:test2} depicts our results. As expected, an increase in $\phi$ increases the zero-fee transactions included and decreases the miner's revenue. These trends are similar to the trend presented in the main paper with Figure~\ref{fig:my_phi}.


\end{document}